\documentstyle[12pt]{article}
\makeatletter
\@addtoreset{equation}{section}
\makeatother

\newtheorem{theorem}{Theorem} 
\newtheorem{corollary}{Corollary} 

\newcommand{\Z}{{\mathchoice{\hbox{$\sf\textstyle Z\kern-0.4em Z$}}
{\hbox{$\sf\textstyle Z\kern-0.4em Z$}}
{\hbox{$\sf\scriptstyle Z\kern-0.3em Z$}}
{\hbox{$\sf\scriptscriptstyle Z\kern-0.3em Z$}}}}
\newcommand{\R}{{\mathchoice{\hbox{$\sf\textstyle I\hspace{-.15em}R$}}     
{\hbox{$\sf\textstyle I\hspace{-.15em}R$}}
{\hbox{$\sf\scriptstyle I\hspace{-.10em}R$}}
{\hbox{$\sf\scriptscriptstyle I\hspace{-.11em}R$}}}}
\newcommand{\C}{{\mathchoice{
\setbox0=\hbox{$\displaystyle\sf C$}\hbox{\hbox
to0pt{\kern0.4\wd0\vrule height0.9\ht0\hss}\box0}}
{\setbox0=\hbox{$\textstyle\sf C$}\hbox{\hbox
to0pt{\kern0.4\wd0\vrule height0.9\ht0\hss}\box0}}
{\setbox0=\hbox{$\scriptstyle\sf C$}\hbox{\hbox
to0pt{\kern0.4\wd0\vrule height0.9\ht0\hss}\box0}}
{\setbox0=\hbox{$\scriptscriptstyle\sf C$}\hbox{\hbox
to0pt{\kern0.4\wd0\vrule height0.9\ht0\hss}\box0}}}}
     
\renewcommand{\Im}{\mbox{\rm Im }}
\renewcommand{\Re}{\mbox{\rm Re }}

\newcommand{\bra}{\langle\,}
\newcommand{\ket}{\,\rangle}

\newcommand{\Tr}{\mbox{\rm Tr}}

\renewcommand{\today}{\number\day \space\ifcase\month\or
  January\or February\or March\or April\or May\or June\or
  July\or August\or September\or October\or November\or December\fi
  \space\number\year}

\newcommand{\ga}{\alpha}
\newcommand{\gb}{\beta}

\newcommand{\gC}{\Gamma}
\newcommand{\gd}{\delta}
\newcommand{\gD}{\Delta}
\newcommand{\gep}{\varepsilon}

\newcommand{\gl}{\lambda}

\newcommand{\gn}{\nabla}

\newcommand{\gO}{\Omega}

\newcommand{\gt}{\theta}
\newcommand{\gT}{\Theta}
\newcommand{\gU}{\Upsilon}
\newcommand{\be}{\begin{eqnarray}}
\newcommand{\ee}{\end{eqnarray}}
\newcommand{\bee}{\begin{eqnarray*}}

\newcommand{\eea}{\end{array} $$}
\newcommand{\eeaa}{\end{array} \ee}
\newcommand{\eee}{\end{eqnarray*}}

\newcommand{\hin}{\mbox{${{\cal H}_{in}}$}}
\newcommand{\hout}{\mbox{${{\cal H}_{out}}$}}

\newcommand{\I}{{\rm i}}
\newcommand{\Int}{\displaystyle \int}
\newcommand{\intoi}{\Int_0^\infty}
\newcommand{\intii}{\Int_{-\infty}^{+\infty}}
\newcommand{\intio}{\Int_{-\infty}^0}

\newcommand{\Lim}{\displaystyle \lim}
\newcommand{\LL}{ {\textstyle L^2(\frac{dp}{2p},\R_+) } } 
\newcommand{\LLk}{ {\textstyle L^2(\frac{dk}{2k},\R_+) } }

\newcommand{\OO}{{\cal O}}
\newcommand{\oo}{{\scriptsize $\bigcirc$}}
\newcommand{\Proof}{{\it Proof.} \ }

\newcommand{\saut}{\hspace{10mm}}

\newcommand{\sch}{{\cal S}(\R)}

\newcommand{\so}{{\cal S}_0(\R)}

\newcommand{\ssp}{{\cal S}(\R_+)}

\newcommand{\Trin}{{\Tr \! \! \! \! \! _{_{_{in}}}} }
\newcommand{\Trout}{{\Tr \! \! \! \! \!  _{_{_{out}}}} }

\newcommand{\vac}{\gO_{o}}

\newcommand{\abs}[1]{\mid \! #1 \! \mid}

\newcommand{\bea}[1]{$$ \begin{array}{#1}}
\newcommand{\beaa}[1]{\be \begin{array}{#1}}
\newcommand{\ch}[1]{{#1
\hspace{-1.65mm}^{\rule[-2.8mm]{0mm}{4.5mm}{\wedge}} }}

\newcommand{\cha}[1]{{#1
\hspace{-2.1mm}^{\rule[-2.8mm]{0mm}{4.5mm}{\wedge} } }}

\newcommand{\fy}[1]{{\ind{#1}{y_o}}}
\newcommand{\fyo}[1]{{{#1}_{y_o}}}
\newcommand{\hc}[1]{{{#1 \!\!}^{^\vee}}}
\newcommand{\ignore}[1]{}
\newcommand{\ind}[2]{{#1 \! \! _{#2}}}
\newcommand{\indice}[2]{{#1 \! _{#2}}}

\newcommand{\LLu}[1]{{ L^1(d #1,\R) }}
\newcommand{\LLum}[1]{{ L^1(d #1,\R_-) }}
\newcommand{\LLup}[1]{{ L^1(d #1,\R_+) }}
\newcommand{\Lu}[1]{{ \parallel \! #1 \!\parallel _{_{L^1}} }}
 
\newcommand{\no}[1]{{ \parallel \! #1 \!\parallel }}

\newcommand{\mtin}[1]{{\bra #1 \ket \! _{_{\gb, in}}}}
\newcommand{\mtout}[1]{{\bra #1 \ket \! _{_{\gb, out}}^{Th}}}

\newcommand{\SS}[2]{ {\rm S}_{#2}{\left[\, #1 \,\right]}}
\newcommand{\sur}[2]{\frac{\textstyle #1}{\textstyle #2}}

\newcommand{\tf}[1]{{{#1 \! \!}^{^\sim}}}
\newcommand{\tfa}[1]{{{#1 \! \! \!}^{^\sim}}}
\newcommand{\tfch}[1]{{\tfa{\ch{#1}}}}

\newcommand{\tfcha}[1]{{\tfa{\cha{#1}}}}

\newcommand{\tfhc}[1]{{\tfa{\rule[0mm]{0mm}{5mm} \hc{#1}}}}
\newcommand{\ul}[1]{\underline{#1}}
\newcommand{\wh}[1]{{\widehat{#1}}}
\newcommand{\sub}[1]{\subsection{}{\vspace{-7.2mm}
{\hspace*{3em} \large\bf #1} 
\\ \\ \hspace{-2mm}}}
\begin{document}
\title{Stimulated emission of particles 
by 1+1 dimensional black holes\footnote{
Work done towards a Ph.D.~at Lausanne University. \\
\mbox{\hspace{4.5mm}}$^\dagger$ 
Present address: Blackett Laboratory,
Theoretical Physics Group, Imperial College, 
London SW7 2BZ, UK.}}
\author{\mbox{}\\ F. Vendrell$^\dagger$ \\ \\
Institut de physique th\'eorique \\
Universit\'e de Lausanne \\
CH-1015 Lausanne, Switzerland\\
E-mail: fvendrel@ipt.unil.ch}
\date{9 October 1996}
\maketitle
\abstract{The stimulated emission of massless bosons by a 
relativistic and the CGHS black hole are studied for real 
and complex scalar fields.
The radiations induced by one-particle and thermal states are 
considered and their thermal properties investigated near the horizon.
These exhibit both thermal and {\it non-thermal} properties for the 
two black-hole models.}
\newpage
\section{Introduction}
The quantum theory of fields in curved space-times is the study of the
propagation of quantum fields in classical gravitational fields
\cite{QFTCST}.
It is a theory of quantum relativity, in the sense that
fields and quantum states look different in distinct non-inertial 
frames \cite{Fu}.
The relativity of the particle concept in curved space-time
implies that particles may be created spontaneously (i.e.~from
the vacuum) for geometrical reasons.
This phenomenon was first analyzed by Parker in cosmology \cite{Pa}
and by Hawking in the gravitational fields of 1+3 dimensional
black holes \cite{Haw}.
It has been shown in particular that the spontaneous emission
of particles is thermal at late times near the event horizon of 
black holes~\cite{BH}.
This discovery subsequently led to the understanding of a 
profound connection between gravity and 
thermodynamics~\cite{Haw76}.

Wald \cite{Wal76} was the first to consider the stimulated 
emission of particles by 1+3 dimensional black holes.
In the presence of an incoming radiation
the emitted radiation must be perturbed somehow or other,
so one may wonder if the emission of particles is still thermal 
near the event horizon of black holes for a non-vacuum incoming 
state.
In the case of the emission of bosons stimulated by 
one-particle states, Wald showed that the mean number of 
particles is indeed still thermal {\it asymptotically} close 
to the event horizon, and that the energy of the incoming 
state must be very large to modify the emission of particles 
close to the event horizon.
In this case, he also observed that the mean number of particles 
emitted is equal to or greater than that for the spontaneous 
emission.
This phenomenon was referred to as {\it amplification} of the 
emitted radiation by Audresch and M\"uller \cite{AM}, who 
confirmed the results obtained by Wald on the basis of a more 
detailed analysis.
In the case of fermions, the emitted radiation is generally 
{\it attenuated} by an incoming state \cite{Fer}.

In this paper I consider the creation of bosons by two 
1+1 dimensional black holes: the relativistic and CGHS black 
holes.
These were introduced in refs \cite{FV} and \cite{CGHS} 
respectively, where some of their semi-classical properties were 
studied.
In particular, for the CGHS black-hole model, it was shown that 
the spontaneous emission of particles is thermal at late times 
near the event horizon.
For the relativistic black-hole model, this emission was shown 
to be thermal everywhere immediately after the formation of 
the black hole.

This paper is specifically concerned with the study of the 
emission of massless bosons stimulated by one-particle and 
thermal states.
Both real and complex scalar fields are considered.
I show that the emitted radiation induced by non-vacuum 
incoming states exhibits both thermal and {\it non-thermal} 
properties near the horizon.
The most remarkable results are the following:
\begin{enumerate}
\item[1)] for both black-hole models and for some one-particle 
incoming states, the two-point function may be non-thermal near 
the horizon;
\item[2)] for both black-hole models and for one-particle and 
thermal incoming states, the energy-momentum tensor of the 
emitted radiation is always thermal close to the horizon;
\item[3)] for the relativistic black-hole model, the mean 
number of particles stimulated by an incoming thermal state 
is {\it not thermal for any test function}.
\end{enumerate}
The cited results of Wald are also confirmed for the two 1+1 
dimensional black-hole models, but I do not use $S$ matrix 
formalism, following Gallay and Wanders \cite{GaW}, since the 
problem is not implementable in these cases \cite{FV,FVTh}.

The second section of this paper is devoted to a review of the
relativistic and CGHS black-hole models.
For later use, the results obtained in ref.~\cite{FV}
on the dynamics of the scalar field in curved space-times are 
summarized in the third section and those for the spontaneous 
emission of bosons in the fourth.
The emission of bosons stimulated by one-particle or thermal 
states is considered in the fifth section.
The mean values of the two-point function, energy-momentum 
tensor, number of created particles and current are computed 
in these states and compared with their corresponding thermal 
mean values.
In the last section the thermal properties of the stimulated 
radiation are discussed for both black-hole models.
\section{Black-hole models in 1+1 dimensions}
The following equation defines a relativistic classical theory of 
gravity in 1+1 dimensional space-times~\cite{Mann}:
\be
R(x) &=& 8\pi G \,T(x),
\label{gravity}
\ee
where $T(x) = T^\mu_{\ \mu}(x)$ is the trace of the energy-momentum 
tensor and $G$ is Newton's constant.
The relativistic black-hole model \cite{FV} is defined assuming
that $T(x)$ is given by
\be
T(x) &=& \sur{M}{8\pi G}\ \gd(x^+-x_o^+),
\label{RBHTEI}
\ee
where $x^\pm=\left(x^0\pm x^1\right)/\sqrt{2}$ and the constant $M$ 
is strictly positive.
Equation (\ref{RBHTEI}) describes a pulse of classical matter.
A solution of eqs (\ref{gravity}) and (\ref{RBHTEI}) is given by
\be
ds^2 &=& \left\{ \begin{array}{lcl}
dx^+\,dx^-, &\hspace{2mm} & \mbox{if \ $x^+< x_o^+$,} \\ [3mm]
\sur{dx^+\,dx^-}{M\,(\gD-x^-)}\,, &&  \mbox{if \ $x^+ > x_o^+$},
\end{array}\right.
\label{RBHmetricx}
\ee
where $\gD$ is an arbitrary constant reflecting the invariance
of the trace (\ref{RBHTEI}) under translations of $x^-$.
The solution (\ref{RBHmetricx}) is not continuous at $x^+=x^+_o$.
Another set of conformal coordinates \mbox{$(y^+,y^-)\in\R^2$} is 
defined by the transformation 
\be
\left\{
\begin{array}{rcl}
x^+(y^+) &=& y^+, \\ [2mm]
x^-(y^-) &=&\gD- e^{-My^-},
\end{array}
\right.
\label{x(y)RBH}
\ee
and in these new coordinates the metric (\ref{RBHmetricx}) is given 
by
\be
ds^2 = \left\{ \begin{array}{ll}
M\,e^{-My^-}\,dy^+dy^-, & \mbox{if \ $y^+< y_o^+$,} \\[2mm]
dy^+dy^-, &  \mbox{if \ $y^+ > y_o^+$,}
\end{array}\right.
\label{RBHmetricy}
\ee
where $y^+_o=x^+_o$.

The CGHS black-hole model \cite{CGHS} is based on the the action
\be
S &=& \sur{1}{2\pi}\, \int d^2x\,\sqrt{-g}\
\left\{\,e^{-2\phi}\,\left[\,R+4\,(\gn \phi)^2+4\,\gl^2 \,\right] \,
-\sur{1}{2}\,(\gn f)^2\,\right\},
\label{CGHSaction}
\ee
where $g$ is the metric, $\phi$ the dilatonic field,
$\gl^2$ the cosmological constant and $f$ a classical matter field.
This action is related to that for non-critical strings
with $c=1$ and defines the dilatonic gravity.
If $T^f_{\mu\nu}(x)$ is the energy-momentum tensor of
the matter field $f$, the CGHS black-hole model is defined by
assuming that
\beaa{rcccl}
T_{++}^f(x) &=& \sur{1}{2}\,(\partial_+ f)^2 
&=& M\,\gd(x^+-x^+_o),
\\ [3mm]
T_{--}^f(x) &=&\sur{1}{2}\,(\partial_- f)^2 &=& 0, \\ [3mm]
T_{+-}^f(x) &=& \sur{1}{2}\,\partial_+ f\,\partial_- f &=& 0,
\label{CGHSTEIf}
\eeaa
where $M>0$.
A continuous solution of the field equations, deduced from the 
action (\ref{CGHSaction}) under the constraints (\ref{CGHSTEIf}), 
is given by
\be
ds^2 &=& 
\left\{ 
\begin{array}{ll}
dx^+dx^-, & \mbox{if \ $x^+< x^+_o$,} \\ [3mm]
\sur{dx^+dx^-}{1+(M/\gl)\,e^{\gl x^-}\,
(\,e^{-\gl x^+}-e^{-\gl x^+_o}\,)}\,, 
\hspace{2mm}& \mbox{if \ $x^+> x^+_o$.}
\end{array}
\right.
\label{CGHSmetricx}
\ee
From this result the scalar curvature is computed:
\be
R(x) &=& 
\left\{ 
\begin{array}{ll}
0, & \mbox{if \ $x^+< x^+_o$,} \\ [3mm]
\sur{4\gl^2}
{1-e^{\gl\,(x^+-x^+_o)}+(\gl/M)\, e^{\gl\,(x^+-x^-)}}\,,
\hspace{2mm}& \mbox{if \ $x^+> x^+_o$.}
\end{array}
\right.
\label{CGHScurvaturex}
\ee
The curvature (\ref{CGHScurvaturex}) is not continuous at 
$x^+=x^+_o$ and it is singular on the curve given by
\be
x^-_{_S}(x^+) &=& x^-_{_H}
-\sur{1}{\gl}\,\log \left[\,1-e^{-\gl\,(x^+-x^+_o)} \,\right],
\saut x^+\in\R,
\label{CGHSsingularity}
\ee
where we have defined
\be
x^-_{_H} &=& x_o^++\sur{1}{\gl}\,\log \sur{\gl}{M}\cdot
\ee
Another set of conformal coordinates \mbox{$(y^+,y^-)\in\R^2$} is 
defined by the transformation 
\be
\left\{
\begin{array}{rcl}
x^+(y^+)&=& y^+, \\ [2mm]
x^-(y^-)&=& x^-_{_H}-\sur{1}{\gl}\,\log 
\left[\,1+\sur{\gl}{M}\,e^{-\gl\,(y^--y_o^+)}\,\right],
\end{array}\right.
\label{x(y)CGHS}
\ee
where $y^+_o=x^+_o$.
In these new coordinates the metric (\ref{CGHSmetricx}) is given by
\be
ds^2 = \left\{ 
\begin{array}{cl}
\sur{dy^+dy^-}{1+(M/\gl)\ e^{\gl (y^--y^+_o)}}\,, 
\hspace{2mm}&  \mbox{if \ $y^+ < y^+_o$,}\\ [4mm]
\sur{dy^+dy^-}{1+(M/\gl)\ e^{\gl (y^--y^+)}}\,, \hspace{2mm}& 
\mbox{if \ $y^+> y^+_o$.}
\end{array}
\right.
\ee
For the two space-time models described, the coordinates 
$x$ and $y$ are Minkowskian if $x^+<x^+_o$ and 
$y^+\rightarrow +\infty$ respectively.
In consequence, these will be called {\it incoming} and 
{\it outgoing} coordinates.
In 1+1 dimensional space-times, I define the {\it horizon}
as the curve on which the conformal factor in the incoming 
coordinates changes its sign, and
the {\it interior} of a black hole as the region of 
space-time for which this conformal factor is negative.
The interior of a black hole is inaccessible to an exterior 
observer and is bounded by the horizon as it should be.
The conformal factor is infinite on the horizon for both 
models and, for the CGHS black-hole model, the horizon also 
coincides with the singularity of the curvature, and so is
given by eq.~(\ref{CGHSsingularity}).
The outgoing coordinates only describe the exterior of the two 
black holes.
\section{Scalar fields in 1+1 dimensional curved space-times}
In ref.~\cite{FV} the dynamics of the massless scalar field was 
studied in the 1+1 dimensional space-times whose incoming and 
outgoing coordinates are related by 
\be
x^\pm &=& x^\pm(y^\pm), \saut y^\pm\in\R,
\ee
as is the case
for the relativistic and CGHS black-hole models.
It was shown that the right and left fields are dynamically 
independent in such space-times, and that it is sufficient to 
consider only the {\it right} moving fields if $x^+(y^+)=y^+$.
The 1+1 dimensional quantum problem is thus reduced to a 
one-dimensional quantum problem, so from now on the suffix
$\pm$ will be dropped.

The incoming and outgoing right fields, denoted by $\phi(x)$ 
and $\ch{\phi}(y)$, were defined in ref.~\cite{FV} from the 
solutions of the massless Klein-Gordon equations in the incoming and 
outgoing coordinates respectively.
They were considered as kernels of distributions.
The set $\so$, on which they act, is the set of the Schwartz 
functions whose Fourier transforms vanish at null momentum, 
\be
\so &=& \{\, f\in\sch \, \mid\,\tf{f}(0)=0\,\},
\ee
so the Wightman distribution is positive definite \cite{MPS}.
The incoming and outgoing field distributions are given by
\beaa{rcccccl}
\phi[h]&=& \intii dx \ h(x)\ \phi(x), &\saut&
\ch{\phi}[f]&=& \intii dy \ f(y)\ \ch{\phi}(y),
\eeaa
if $h,f\in\so$.

The {\it incoming test function} $\ch{f}(x)$ was defined in 
ref.~\cite{FV} in terms of the {\it outgoing test function} $f(y)$ 
through the scalar properties of the field distributions:
\be
\phi[\ch{f}] &=& \ch{\phi}[f].
\label{transfField}
\ee
The transformation of the test functions describes the propagation 
of the quantum fields.
From eq.~(\ref{transfField}), these are related by the 
operator $U$, as $\tfch{f}=U\tf{f}$, whose kernel is given 
by
\be
U(k,p)&=& \sur{1}{2\pi} \intii dy\ e^{-\I kx(y)}\ e^{\I py},
\ee
where $k$ and $p$ are the {\it incoming} and {\it outgoing 
momentum} respectively.
For the relativistic and black-hole models, this is given by 
\cite{FV,CGHS} 
\be
U_{_R}(k,p) &=& 
\sur{e^{-\I\,k\gD}\,e^{-\I\,\gO\left(\frac{p}{M}\right)}\,
e^{\I \frac{p}{M} \log \vert k \vert}}{\sqrt{2\pi\,M}}\ 
\left[ \ \sur{\gt(k)}{\sqrt{p\,(1-e^{-\frac{2\pi}{M} p})}}
+\sur{\gt(-k)}{\sqrt{p\,(e^{\frac{2\pi}{M}p}-1)}} \ \right],
\label{UkernelTNR} \\ [2mm]
U_{_{CGHS}}(k,p) &=& \sur{e^{\I\,x_o (p-k)}}{2\pi\gl}\, 
\left(\sur{\gl}{M}\right)^{\I\,\frac{(p-k)}{\gl}}\,
B\left(\,\I\,\frac{(p-k)}{\gl}+0^+,-\I\,\sur{p}{\gl}+0^+\,\right),
\label{UkernelCGHS}
\ee
where $\gO(p)= \arg \left[\,\gC(ip)\,\right]$ and $B$ is the beta 
function.

The {\it incoming} and {\it outgoing Hilbert spaces}, $\hin$ and 
$\hout$, were defined using the incoming and outgoing fields.
The {\it incoming} and {\it outgoing wave function spaces} are 
given by $\LLk$ and $\LL$ respectively.

In ref.~\cite{FV} the mean values of several observables, 
constructed in the outgoing coordinates, were also computed in the 
incoming vacuum.
These describe the properties of the outgoing radiation in the exterior
of the black holes.
The two-point function, energy-momentum tensor, number
of created particles and current were considered. 
I repeat here their definitions and the results obtained, since they 
will be useful below.
They are valid for both real and complex scalar fields.
\\ \\
$\bullet$ Incoming and outgoing two-point functions:
\beaa{rcccl}
W_o(x,x') 
&=& (\vac,\phi(x)\,\phi(x')^\dagger\,\vac)
&=& - \sur{1}{4\pi}\ \log \left[\,x'-x+\I\,0^+ \,\right],
\hspace{10mm} \label{Sp2PFx}
\eeaa
\mbox{} \vspace{-13mm} \\
\beaa{rcccl}
\wh{W}_o(y,y') &=& (\vac,\ch{\phi}(y)\,\ch{\phi}(y')^\dagger\,\vac)
&=& - \sur{1}{4\pi}\ \log \left[\,x(y')-x(y)+\I\,0^+ \,\right].
\label{Sp2PFy}
\eeaa
Equations involving two-point functions are always and only valid
between kernels of distributions on \mbox{$\so\times\so$}.
In particular,
\be
(\vac, \phi[g_1]\,\phi[g_2]^\dagger\,\vac) &=&
\intoi \sur{dk}{2k} \ \tfa{g}_2(k)^* \,\tfa{g}_1(k),
\label{one2pf}
\ee
where $\tfa{g}_1,\tfa{g}_2\in\so$.
\\ \vspace{-3mm} \\
$\bullet$ Energy-momentum tensor:
\beaa{rclcl}
\wh{T}_o(y) &=& (\vac ,:\wh{\gT}\, (y) :_{out} \, \vac) \\ [3mm]
 &=&  \Lim_{\gep \rightarrow 0}\
(\vac, \left[ \,\wh{\gT}_\gep (y)\, -
\gT_\gep (x(y)) \, \right]\, \vac)
&=& -\sur{1}{24 \pi} \, \SS{x(y)}{y},
\label{SpTEI} 
\eeaa
where $\mbox{S}_y$ is the Schwartz derivative and
\be
\gT(x) &=& \partial_x \phi(x)^\dagger\,\partial_x \phi(x),
\label{TEIopx} \\ [2mm]
\wh{\gT}(y) &=& 
\partial_y \ch{\phi}(y)^\dagger\,\partial_y \ch{\phi}(y),
\label{TEIopy} \\ [3mm]
\gT_\gep (x) &=&\frac{1}{2} \left[ \,
\partial_x \phi(x)^\dagger \, \partial_x \phi(x+\gep) +
\partial_x \phi(x+\gep)^\dagger \, \partial_x \phi(x) \, \right], 
\label{TEIxbis}\\ [3mm]
\wh{\gT}_\gep (y) &=& \frac{1}{2} \left[ \,
\partial_y \ch{\phi}(y)^\dagger\,\partial_y \ch{\phi}(y+\gep) +
\partial_y \ch{\phi}(y+\gep)^\dagger\,
\partial_y \ch{\phi}(y)\,\right].
\label{TEIybis}
\ee
$\bullet$ Mean number of spontaneously created particles:
\beaa{rclcl}
\bar{N}_o[f] &=& 
(\vac, \ch{\phi}[f]^\dagger\,\ch{\phi}[f]\,\vac)
&=& \intoi \sur{dk}{2k} \left\vert (U\tf{f})(-k)\right\vert^2,
\label{number}
\eeaa
if $\tf{f}\in\ssp$ is a normalized test function, where
$\ssp$ is set of Schwartz functions whose Fourier transforms
vanish for negative momentum.
\\ \vspace{-3mm} \\
$\bullet$ Mean current:
\beaa{rclcl}
\wh{J}_o(y) &=& (\vac, :\wh{\gU}(y):_{out} \, \vac )
\\ [3mm]
&=& \Lim_{\gep \rightarrow 0} \ (\vac,
\left[ \, \wh{\gU}_\gep (y) -\gU_\gep (x(y))\,\right]
\, \vac )
&=& 0,
\label{current}
\eeaa
where
\be
\gU (x) &=& \I \ \phi(x)^\dagger
\stackrel{\leftrightarrow}{\partial}_x \phi(x),
\label{currentx} \\ [3mm]
\wh{\gU} (y) &=& \I \ \ch{\phi}(y)^\dagger
\stackrel{\leftrightarrow}{\partial}_y \ch{\phi}(y),
\label{currenty} \\ [3mm]
\gU_\gep (x)&=& \I\,\left[\
\phi(x+\gep)^\dagger \, \partial_x \phi(x)
- \partial_x \phi(x)^\dagger\,. \phi(x+\gep)\ \right],
\label{currentxbis} \\ [2mm]
\wh{\gU}_\gep (y)&=& \I\,\left[\
\ch{\phi}(y+\gep)^\dagger \, \partial_y \ch{\phi}(y)
- \partial_y \ch{\phi}(y)^\dagger\,. 
\ch{\phi}(y+\gep)\ \right].
\label{currentybis}
\ee

The {\it thermal mean value} of an observable $A$ in the 
outgoing Hilbert space $\hout$ was defined in ref.~\cite{FV} 
by\footnote{Note that it has to be defined as a limit.}
\be
\mtout{A} &=&
\sur{\Trout\left[\,e^{-\gb H_{out}}\,A\,\right]}
{\Trout\left[\,e^{-\gb H_{out}}\,\right]}\,,
\label{ThermalMV}
\ee
if the representation of $A$ in $\hout$ is known.
For the above observables
\be
W_{\gb,out}^{Th}\, (y,y') &=&
- \sur{1}{4\pi} \log \left\{\, \sur{\gb}{\pi} \,
\sinh \left[\,\frac{\pi}{\gb}\,\left(y'-y+\I\,0^+\right) 
\,\right]\,\right\},
\label{2PFthermal} \\ [2mm]
T_{\gb,out}^{Th}(y) 
&=&T_{\gb,out}^{Th}
\ = \ \sur{\pi}{12 \gb^2}\,, \saut \forall\, y\in\R,
\label{thermalTEI} \\ [2mm]
\bar{N}_{\gb,out}^{Th}[f]
&=& \int_0^\infty \sur{dp}{2p} 
\sur{\abs{\tf{f}(p)}^2}{e^{\gb p}-1}\,,
\label{Nthermal} \\ [2mm]
J^{Th}_{\gb,out}(y) &=& 0,  \saut \forall\, y\in\R.
\label{thermalCurrent}
\ee
It has been assumed in eq.~(\ref{Nthermal}) that $\tf{f}\in\LL$ is 
normalized and that $f(y)$ exists a.e.~and is integrable.
\section{Spontaneous creation of particles}
Using the results of the preceding section 
with the relativistic black-hole model \cite{FV}
and transformation (\ref{x(y)RBH}) gives
\be
\wh{W}_o\,(y,y') &=& W_{\frac{2\pi}{M},out}^{Th}\,(y,y'), 
\saut \forall\,y,y'\in\R,
\label{Sp2PFRBH}
\\ [3mm]
\wh{T}_o(y) &=& T_{\frac{2\pi}{M},out}^{Th}\,, 
\saut \forall\,y\in\R,
\label{SpTEIRBH} \\ [3mm]
\bar{N}_o[f] &=& \bar{N}_{\frac{2\pi}{M},out}^{Th}\,[f],
\label{SpNRBH} \\ [2mm]
\wh{J}_o(y) &=& 0,  \saut \forall\, y\in\R.
\label{SpCurrentRBH}
\ee
It has been assumed in eq.~(\ref{SpNRBH}) that $\tf{f}\in\LL$ is 
normalized and that $f(y)$ exists a.e.~and is integrable.
From these results we conclude that the emission of particles is 
thermal everywhere after the formation of the black hole and that the 
associated temperature is given by $T=\frac{M}{2\pi}$. 

Similarly, for the CGHS black-hole model \cite{FVTh},
and transformation (\ref{x(y)CGHS}),
\be
\wh{W}_o \, (y,y') &\approx&
W_{\frac{2\pi}{\gl},out}^{Th}\, (y,y'),
\saut \mbox{if \ $y,y'\gg 1$,}
\label{Sp2PFCGHSa} \\ [2mm]
\wh{W}_o (y,y') &\approx & W_{\infty,out}^{Th} (y,y'),
\saut \hbox{if \ $-y, \, -y' \gg 1$,}
\label{Sp2PFCGHSb} \\ [2mm]
\lim_{y\rightarrow +\infty}
\wh{T}_o (y) &=& T^{Th}_{\frac{2\pi}{\gl},out}\,,
\label{SpTEICGHSa} \\ [2mm]
\lim_{y\rightarrow -\infty}
\wh{T}_o (y) &=& T^{Th}_{\infty,out}\,,
\label{SpTEICGHSb} \\ [2mm]
\wh{J}_o(y) &=& 0,  \saut \forall\, y\in\R.
\label{SpCurrentCGHS}
\ee
These results suggest that the emission of particles is 
thermal near the horizon at late times and that the 
associated temperature is given by $T=\frac{\gl}{2\pi}$.
However the situation turns out to be more complex after considering 
the mean number $\bar{N}_o[f]$ of spontaneously created particles.

Asymptotically close to the horizon, the behavior of  
$\bar{N}_o[f]$ is analyzed in the following way.
The translation of the test function $f$ towards the 
horizon \cite{Wan} is defined as
\be
f_{y_o}(y) &=& f(y-y_o),
\label{deffyo}
\ee
and we examine whether
\be
\lim_{y_o \rightarrow +\infty} 
\left( \, \bar{N}_o[f_{y_o}] -
\bar{N}_{\frac{2\pi}{\gl},out}^{Th}[f_{y_o}] \, \right) 
&\stackrel{?}{=}& 0.
\label{SNthermal}
\ee
\begin{theorem} \label{th:SpCloseN}
Let $\tf{f}\in\LL$ be a normalized wave function such that $f(y)$
exists a.e.~and is integrable.
If there exist three constants \ul{$\ga > 1/2$}, $C>0$ and
$L \geq 1$ such that
\be
\abs{f(y)} &\leq& \sur{C}{\abs{y}^{1+\ga}}\,,
\saut \mbox{if \ $y \, \leq - L $,}
\label{SpalgebraicA} 
\ee 
then for the CGHS model and for all $y_o > 0$,
\beaa{rcl}
\left\vert \, \bar{N}_o[f_{y_o}] -
\bar{N}^{Th}_{\frac{2\pi}{\gl},out}[f_{y_o}] \, \right\vert
&\leq &
\sur{32\,C^2}
{\ga^2(2\ga-1)\left(\,\frac{1}{4}\,y_o+L-1\,\right)^{2\ga-1}}
\\ [7mm] &&
+ \ e^{2L-y_o/2} \left(\, \Lu{f}+ \Lu{f'} \right)^2;
\label{SpNCGHS}
\eeaa
however if
\be
f(y) &\approx& \sur{C}{(-y)^{1+\ga}}\,,
\saut \mbox{if \ $y \, \ll -1 $,}
\label{SpalgebraicB} 
\ee 
where \ul{$0<\ga\leq 1/2$} and $C\in\C$, then
\be
\bar{N}_o[f_{y_o}] \ =\
\bar{N}_{\frac{2\pi}{\gl},out}^{Th}[f_{y_o}] &=& \infty
\saut \mbox{$\forall\,y_o\in\R$,} \\ [2mm]
\left\vert \, \bar{N}_o[f_{y_o}] -
\bar{N}_{\frac{2\pi}{\gl},out}^{Th}[f_{y_o}] \, \right\vert 
&=& \infty,
\saut \mbox{$\forall\,y_o\in\R$.} 
\label{SpNotThN}
\ee 
\end{theorem} 
This theorem is proved in ref.~\cite{FVTh}.
Under some conditions, the bound (\ref{SpNCGHS}) shows that 
eq.~(\ref{SNthermal}) is true if the modulus of $f$ decreases 
sufficiently fast {\it very far from the horizon}.
This bound is composed of two terms:
the first decreases algebraically in $y_o$ with exponent
$2\ga-1$, and the second decreases exponentially in $y_o$.
However, if the function $f$ decreases relatively weakly 
{\it very far from the horizon} and does not oscillate, 
eq.~(\ref{SpNotThN}) shows that eq.~(\ref{SNthermal}) is not 
true.
This result implies that {\it the behavior of some non-local 
observables may be non-thermal near the horizon}.

There exists a bound for $\bar{N}_o[f]$ in terms of the Fourier 
transform $\tf{f}$.
\begin{theorem} \label{th:SpBoundN}
If $\tf{f}\in\LL$ is a normalized wave function such that $f(y)$
exists a.e.~and is integrable, then for the CGHS model
\be
\bar{N}_o[f] &\leq& C \Int_0^\infty \sur{dp}{2p} \, 
\sur{\abs{\tf{f}(p)}^2}{1-e^{-\frac{2\pi}{\gl} p}}\,,
\label{SpBoundN}
\ee 
where $C>0$ is a constant.
\end{theorem}
This theorem is also proved in ref.~\cite{FVTh}.
The bound (\ref{SpBoundN}) may be infinite depending on the infrared 
behavior of $\tf{f}$.
\section{Emission stimulated by a one-particle state}
For simplicity I will assume from now on that $M=\gl=1$
and $\gD=x^+_o=0$.
In this case, the transformation between the (right) incoming
and outgoing coordinates is given for the relativistic
black-hole model by
\be
x(y) &=& -e^{-y},
\saut \forall\,y\in\R,
\ee
and for the CGHS model by
\be
x(y) &=& -\log \left(1+e^{-y}\right),
\saut \forall\,y\in\R,
\ee
see eqs (\ref{x(y)RBH}) and (\ref{x(y)CGHS}).
Both transformations apply $\R$ on $\R_-$ and the first is the
asymptotical form of the second for $y \gg 1$.

In this section we assume that the incoming state is a
one-particle state $\gO_g\in\hin$.
For the real scalar field this is defined as
\be
\gO_g &=& \phi[g]^\dagger\,\vac,
\ee
where $\tf{g}\in\ssp$ and $\vac\in\hin$ is the incoming 
vacuum.
For the complex scalar field, the one-antiparticle state 
is also defined as
\beaa{rcccl}
\bar{\gO}_g &=& \gO_{g^*} &=&\phi[g^*]^\dagger\,\vac,
\eeaa
where $\tf{g}\in\ssp$.
The first subsections of this section are devoted to the real 
scalar field and the last one to the complex scalar field.
\subsection{The two-point function}
For the incoming state $\gO_g$, the incoming and outgoing 
two-point functions denoted by $W_g(x,x')$ and $\wh{W}_g(y,y')$ 
respectively, are defined as
\be
W_g(x,x') &=& (\gO_g, \phi(x)\, \phi(x')^\dagger \, \gO_g), 
\label{Sg2PFx} \\ [2mm]
\wh{W}_g (y,y') &=& 
(\gO_g, \ch{\phi}(y)\,\ch{\phi}(y')^\dagger\,\gO_g),
\label{Sg2PFy}
\ee
and are related by
\be
\wh{W}_g (y,y') &=& W_g (x(y),x(y')),
\label{1P2Pfxy}
\ee
since $\ch{\phi}(y)=\phi(x(y))$ for all $y\in\R$ \cite{FV}.
These functions may be expressed in terms of the 
primitive $G$ of $g$ defined as
\be
G(x) &=& \int_{-\infty}^x dx'\,g(x').
\label{defG}
\ee
\begin{theorem} \label{th:Sg2PF}
Between kernels of distributions on $\so\times\so$,
\be
W_g (x,x') &=& W_o (x,x')+\sur{1}{2} \,\Re [\,G(x) \, G(x')^*\,],
\label{S2PFx} \\ [2mm]
\wh{W}_g (y,y') &=& \wh{W}_o (y,y') + 
\sur{1}{2} \, \Re [\,G(x(y)) \, G(x(y'))^*\,],
\label{S2PFy}
\ee
where $W_o (x,x')$ and $\wh{W}_o (y,y')$ are the two-point 
functions (\ref{Sp2PFx}) and (\ref{Sp2PFy}).
\end{theorem}
\Proof
The two-point function (\ref{Sg2PFx}) is computed using Wick's 
theorem and
\be
(\vac, \phi[g]\,\phi(x)\,\vac) &=& \sur{\I}{2}\,G(x),
\ee
from which eq.~(\ref{S2PFx}) is obtained.
Equation (\ref{S2PFy}) follows from the property (\ref{1P2Pfxy}).
\hfill{$\Box$} \\ \\
This last theorem applies to both black-hole models.
\begin{theorem} \label{th:2PFBH}
Let $\tf{g}\in\ssp$ be an incoming normalized test function.
For the relativistic and CGHS models, we have the following 
asymptotic relation between kernels of distributions on 
\mbox{$\so\times\so$}, 
\beaa{rccccccc}
G(0) &=& 0 &\Longleftrightarrow&
\wh{W}_g (y,y') &\approx& W_{2\pi,out}^{Th}(y,y') \ \
&\mbox{if \ $y,\, y'\gg 1$};
\label{S2PFthA}
\eeaa
furthermore, for the relativistic model,
\be
\wh{W}_g (y,y') &\approx& W_{2\pi,out}^{Th}(y,y'), \saut
\mbox{if \ $-y,\, -y'\gg 1$},
\label{S2PFthB}
\ee
and for the CGHS model,
\be
\wh{W}_g (y,y') &\approx & W_{\infty,out}^{Th} (y,y'),
\saut \hbox{if \ $-y, \, -y' \gg 1$,}
\label{S2PFthC}
\ee
where $W_{\gb,out}^{Th}(y,y')$ is given by 
eq.~(\ref{2PFthermal}).
\end{theorem}
\Proof
For both models, we have $x(y)\approx 0$ if $y\gg 1$.
Equation (\ref{S2PFy}) implies that
\be
\wh{W}_g (y,y') &\approx & 
\wh{W}_o (y,y') + \sur{1}{2}\,\abs{G(0)}^2,
\saut \mbox{if \ $y, \, y' \gg 1$,} 
\label{eq:stimuleWgA}
\ee
since $G$ is continuous.
Equation (\ref{S2PFthA}) is then deduced from the results 
(\ref{Sp2PFRBH}) and (\ref{Sp2PFCGHSa}).
Equation (\ref{S2PFy}) also implies that
\be
\wh{W}_g (y,y') &\approx & \wh{W}_o (y,y'),
\saut \mbox{if \ $-y, \, -y' \gg 1$},
\ee
since $G$ vanishes at infinity.
Equations (\ref{S2PFthB}) and (\ref{S2PFthC}) are deduced from 
the results (\ref{Sp2PFRBH}) and (\ref{Sp2PFCGHSb}).
\hfill{$\Box$} \\ \\
For both black-hole models, this theorem states that, in the 
outgoing coordinates, the incoming state $\gO_g$ is thermal 
\cite{FV} close to the horizon {\it if and only if} $G(0)=0$, 
and in this case the temperature is given by $(2\pi)^{-1}$.
Very far from the horizon, the incoming state $\gO_g$ is also 
thermal and the associated temperature is given by $(2\pi)^{-1}$ 
for the relativistic model and vanishes for the CGHS model.
\subsection{The energy-momentum tensor}
The energy-momentum observables in the incoming and outgoing
coordinates are given respectively by eqs (\ref{TEIopx}) and 
(\ref{TEIopy}).
Their regularized mean values in the incoming state $\gO_g$ will 
be denoted by $T_g(x)$ and $\wh{T}_g(y)$ respectively.
In the normal and subtraction regularization schemes, these are 
given by
\be
T_g(x) &=& (\gO_g, :\gT(x):_{in} \, \gO_g), 
\label{defgTEInormalx} \\ [2mm]
T_g(x) &=& \lim_{\gep\rightarrow 0} \, 
\left[ \, (\gO_g, \gT_\gep (x) \, \gO_g)
- (\vac,\gT_\gep (x) \, \vac) \, \right], 
\label{defgTEIsubx}\\ [2mm]
\wh{T}_g(y) &=& (\gO_g, :\wh{\gT}(y):_{out} \, \gO_g),
\label{defgTEInormaly} \\ [2mm]
\wh{T}_g(y) &=& \lim_{\gep\rightarrow 0} \, 
\left[ \, (\gO_g, \wh{\gT}_\gep (y) \, \gO_g)
- (\vac,\gT_\gep (x(y)) \, \vac) \, \right],
\label{defgTEIsuby}
\ee
where $\gT_\gep (x)$ and $\wh{\gT}_\gep (y)$ are defined by eqs
(\ref{TEIxbis}) and (\ref{TEIybis}) respectively.
\begin{theorem} \label{th:simpleTEI}
If $\tf{g}\in\ssp$ is a normalized test function, 
and if $T_g(x)$ and $\wh{T}_g(y)$ are defined respectively by
eqs (\ref{defgTEInormalx}) and (\ref{defgTEInormaly}), or by
eqs (\ref{defgTEIsubx}) and (\ref{defgTEIsuby}), then
\be
T_g(x) &=& \frac{1}{2}\, \abs{g(x)}^2, 
\label{gTEIx}\\ [2mm]
\wh{T}_g(y) &=& \wh{T}_o(y)+ x'(y)^2\ T_g(x(y)),
\label{gTEIy}
\ee
where $\wh{T}_o(y)$ is given by eq.~(\ref{SpTEI}).
\end{theorem}
This theorem is proved is appendix \ref{ap:simpleTEI}.
It applies to both black-hole models.
\begin{theorem} \label{th:simpleTEIBH}
If $\tf{g}\in\ssp$ is a normalized incoming test function,
then for both models
\be
\lim_{y\rightarrow +\infty} \wh{T}_g(y) 
&=& T_{2\pi,out}^{Th}\,;
\label{simpleTEIA}
\ee
furthermore for the CGHS model
\be
\lim_{y\rightarrow -\infty} \wh{T}_g(y) 
&=& T_{\infty,out}^{Th}\,,
\label{simpleTEIB}
\ee
where $T_{\gb,out}^{Th}$ is given by eq. 
(\ref{thermalTEI}).
\end{theorem}
\Proof
For both models, we have $x'(y)\approx e^{-y}$ if 
$y\gg 1$.
Equation (\ref{gTEIy}) implies that
\be
\wh{T}_g(y) - \wh{T}_o(y) &\approx& 
\sur{1}{2}\,e^{-2y} \, \abs{g(0)}^2, \saut \mbox{if \ $y\gg 1$,}
\ee
from which eq.~(\ref{simpleTEIA}) is deduced using the results 
(\ref{SpTEIRBH}) and (\ref{SpTEICGHSa}).
For the CGHS model, we have $x'(y)\approx 1$ if 
$-y\gg1$.
Since $g$ vanishes at infinity, eq.~(\ref{gTEIy}) also implies
that
\beaa{rcccl}
\wh{T}_g(y) - \wh{T}_o(y) &\approx&
\sur{1}{2}\, \abs{g(y)}^2  &\approx& 0, \saut 
\mbox{if \ $-y\gg 1$,}
\eeaa
from which eq.~(\ref{simpleTEIB}) is deduced using the result
(\ref{SpTEICGHSb}).
\hfill{$\Box$} \\ \\
For both black-hole models, the behavior of the energy-momentum
tensor is thus still thermal close to the horizon for the 
incoming one-particle state $\gO_g$, {\it even if the test 
function $g$ is well localized in the vicinity of the horizon}.
Furthermore, in the case of the CGHS model, an 
incoming one-particle state does not induce any radiation very 
far from the horizon.

In a given set of coordinates, the total energy is given by the 
integral of the energy-momentum tensor.
The incoming and outgoing energies, $E^{in}$ and $E^{out}$, are
defined as
\beaa{rcccccl}
E^{in} &=& \intii dx \ T(x), &\saut&
E^{out} &=& \intii dy \ \wh{T}(y).
\label{energies}
\eeaa
The energy $E^{out}_g$ of the radiation stimulated by the 
one-particle state $\gO_g$ is always larger than the energy 
$E^{out}_o$ of the spontaneously emitted radiation, as shown in 
the theorem below.
This theorem gives a necessary condition for the outgoing 
energy $E^{out}_g$ to be smaller than the sum of the incoming 
energy $E^{in}_g$ of the one-particle state $\gO_g$ and the 
spontaneously emitted energy $E^{out}_o$.
\begin{theorem} \label{th:simpleenergies}
If $\tf{g}\in\ssp$ is a normalized incoming test function, then
for any transformation of coordinates $x=x(y)$,
\be
E_o^{out} &<& E_g^{out}.
\label{INenergiesA}
\ee
Moreover if
\be
x'(y) &<& 1, \saut \forall\,y\in\R,
\label{Hx'(y)}
\ee
then
\be
E_g^{out} &<& E_o^{out}+ E^{in}_g.
\label{INenergiesB}
\ee
\end{theorem}
\Proof
Together with theorem \ref{th:simpleTEI}, definitions 
(\ref{energies}) imply that
\be
E_g^{out}-E_o^{out} &=& 
\frac{1}{2} \intii dy \ x'(y)^2 \, \abs{g(x(y))}^2,
\ee
from which eq.~(\ref{INenergiesA}) is deduced.
Since
\be
E^{in}_g &=& \frac{1}{2} \intii dx \, \abs{g(x)}^2,
\ee
eq.~(\ref{INenergiesB}) is obtained under the hypothesis 
(\ref{Hx'(y)}).
\hfill{$\Box$} \\ \\
Hypothesis (\ref{Hx'(y)}) is not satisfied by the
relativistic model but by the CGHS model.
For this last model only, eq.~(\ref{INenergiesB}) shows that 
part of the incoming energy is absorbed by the black hole.
\subsection{The mean number of created particles}
\label{sub:MNCP}
The mean number $\bar{N}_g[f]$ of particles stimulated by the 
incoming state $\gO_g$ is given by
\be
\bar{N}_g[f] 
&=& (\gO_g,\, \ch{\phi}[f]^\dagger\,\ch{\phi}[f]\,\gO_g),
\ee
where $\tf{f}\in\ssp$.
The operators $A$ and $B$ are defined respectively as the 
positive and negative contributions of $U$ to the incoming 
momentum:
\beaa{rclcrcl}
(A\tf{f})(k) &=& \gt(k)\,(U\tf{f})(k), &\saut& 
(B\tf{f})(k) &=& \gt(k)\,(U\tf{f})(-k).
\label{defAB}
\eeaa
\begin{theorem} \label{th:SN}
If $\tf{f}, \tf{g}\in\ssp$ are two normalized test functions, 
then for any transformation of coordinates $x(y)$,\footnote{
Equation (\ref{SN}) was first obtained by Wald \cite{Wal76}.
}
\be
\bar{N}_g[f] &=& \bar{N}_o[f] + 
\left\vert\,\bra\tf{g},A \tf{f}\ket\,\right\vert^2 + 
\left\vert\,\bra \tf{g}^*, B\tf{f}\ket\,\right\vert^2.
\label{SN}
\ee 
\end{theorem}
\Proof
This theorem is proved using the field 
transformation (\ref{transfField}) and Wick's theorem.
\hfill{$\Box$} \\ \\
Equation (\ref{SN}) may be extended by continuity to the wave 
functions $\tf{f}\in\LL$ and \mbox{$\tf{g}\in\LLk$} if $f(y)$ 
and $g(x)$ exist a.e.~and are integrable.

Some general bounds for $\bar{N}_g[f]$ are now given.
The classical incoming function $\indice{\tf{g}}{cl}$ has been
defined in ref.~\cite{Wal76} in terms of the outgoing test 
function $f(y)$ by
\be
\indice{\tf{g}}{cl}(k) &=& \no{A\tf{f}}^{-1}\,(A\tf{f})(k).
\label{gclassical}
\ee
\begin{theorem}
If $\tf{f}\in\LL$ and $\tf{g}\in\LLk$ are two normalized wave 
functions such that $f(y)$ and $g(x)$ exist a.e.~and are 
integrable, then for any transformation of coordinates 
$x(y)$,\footnote{
Equation (\ref{SBoundB}) was first obtained by Wald \cite{Wal76}.
}
\beaa{rcccl}
\bar{N}_o [f] &\leq& \bar{N}_g [f] &\leq& 1+3 \, \bar{N}_o[f], 
\label{SBoundA}
\eeaa
\mbox{}\vspace{-13mm}\\
\beaa{rcccl}
1+ \bar{N}_o[f] &\leq& \bar{N}_{g_{cl}}[f]. 
&\mbox{\hspace{5mm}}&
\label{SBoundB}
\eeaa
\end{theorem}
\Proof
To obtain eq.~(\ref{SBoundA}), the Cauchy-Schwartz inequality 
is applied to eq.~(\ref{SN}) using the fundamental relation 
$A^\dagger\,A = B^\dagger\, B+E$, where E is the 
identity \cite{FV}. 
The inequality (\ref{SBoundB}) is deduced from 
eq.~(\ref{SN}) and def.~(\ref{gclassical}).
\mbox{\ }\hfill{$\Box$} \\ \\
This theorem states that $\bar{N}_g[f]$ is finite if and only 
if $\bar{N}_o[f]$ is finite, and that $\bar{N}_g[f]$ is always 
equal to or larger than $\bar{N}_o[f]$.
For the classical incoming function it implies, in this 
particular case, that the mean number of created particles 
stimulated by the state $\gO_g$ is larger than the sum of the 
mean number of incoming particles and the mean number of 
spontaneously created particles, although the opposite inequality 
(\ref{INenergiesB}) is satisfied for the incoming and outgoing 
energies.

For both black-hole models, bounds for $\bar{N}_g [f]$ and 
$\bar{N}_g[f]-\bar{N}_o[f]$ exist in terms of the Fourier 
transform $\tf{f}$.
\begin{theorem} \label{th:SgBound}
If $\tf{f}\in\LL$ and $\tf{g}\in\LLk$ are two normalized wave 
functions such that $f(y)$ and $g(x)$ exist a.e.~and are 
integrable, then for both models
\be
\bar{N}_g[f] &\leq&
C\, \intoi \sur{dp}{2p} \, 
\sur{\abs{\tf{f}(p)}^2}{1-e^{-2\pi p}}\,,
\label{SgBoundN}
\ee
where $C>0$ is a constant which does not depend on the 
incoming function $g$.
\end{theorem}
\Proof
Equation (\ref{SgBoundN}) follows from eq.~(\ref{SBoundA})
using the result (\ref{SpNRBH}) and theorem \ref{th:SpBoundN}.
\mbox{}\hfill{$\Box$}
\begin{theorem} \label{th:Sbound}
If $\tf{f}\in\LL$ and $\tf{g}\in\LLk$ are two normalized wave
functions such that $f(y)$ and $g(x)$ exist a.e.~and are 
integrable, then for both models,
\be
\left\vert\,\bar{N}_g[f]-\bar{N}_o[f]\,\right\vert &\leq &
\sur{1}{\pi}\ \no{g}_{\LLum{x}}^2 \, 
\left(\,\intoi \sur{dp}{2p} \, \abs{\tf{f}(p)}\,\right)^2.
\label{Sbound}
\ee
\end{theorem}
This theorem is proved in appendix \ref{ap:Sbound}.
It shows in particular that the mean number of created extra 
particles $\bar{N}_g[f]-\bar{N}_o[f]$ may be finite even if 
$\bar{N}_o[f]$ and $\bar{N}_g[f]$ are both infinite. 
The bound (\ref{Sbound}) depends only on the norm of 
$g$ in $\LLum{x}$, i.e.~only on the restriction of $g$ to the 
{\it exterior} of the black holes. 

The outgoing function $f_{p_o}$ of mode $p_o>0$ is defined 
as\footnote{The definition of the null mode $f_{p=0}$ is given 
in ref.~\cite{FV}.}
\be
\ind{\tf{f}}{{p_o}}(p) &=& 2p\ \gd(p-p_o),
\saut \forall\,p\geq 0,
\label{Sdefmode}
\ee
and the outgoing function $\hc{g}(y)$ is defined in terms of 
the incoming function $g(x)$ as
\be
\hc{g}(y) &=& x'(y)\, g(x(y)), \saut \forall\ y\in\R.
\label{Sdefgoutgoing}
\ee
There is a bound for the mean number of particles in a given 
outgoing mode.
The total mean number of particles, which is defined as 
\cite{FV}
\be
\bar{N}^{tot} &=& \intoi \sur{dp}{2p} \ \bar{N}[f_p],
\label{Ntot} 
\ee
can also computed for a one-particle incoming state.
\begin{theorem} \label{th:SNmode}
If $\tf{g}\in\LLk$ is a normalized incoming test function, then 
for both black holes\footnote{
Equation (\ref{SNtotA}) is in fact true for all transformations
$x=x(y)$.} 
and for $p\geq 0$,
\be
\bar{N}_g[f_p] &\leq& \bar{N}_o[f_p] 
+ \sur{1}{\pi}\ \no{g}_{\LLum{x}}^2\,,
\label{SNmode} \\ [2mm]
\bar{N}_g^{tot} &=& \bar{N}_o^{tot} + \no{\tfhc{g}}^2,
\label{SNtotA} 
\ee
and
\be
\bar{N}_g^{tot} &=& \infty;
\label{SNtotB}
\ee
moreover if $\tf{g}\in\ssp$, the difference 
$\bar{N}_g^{tot}-\bar{N}_o^{tot}$ is finite if and only if 
$G(0)=0$, where $G$ is defined by eq.~(\ref{defG}).
\end{theorem}
This theorem is proved in appendix \ref{ap:SNmode}.
For both black-hole models, it states that the incoming state 
$\gO_g$ induces a finite mean number of extra particles in a 
given mode.
The {\it total} mean number of extra particles created may be 
finite or infinite depending on $g$.
\subsection{Close to the horizon}
The behavior of the mean number $\bar{N}_g[f]$ of stimulated 
particles is now examined close to the horizon by computing
a bound for $\bar{N}_g[f_{y_o}] -\bar{N}_o[f_{y_o}]$ in terms of
$y_o$, where the function $f_{y_o}$ is defined in 
eq.~(\ref{deffyo}).
\begin{theorem} \label{th:ScloseN}
Let $\tf{f}\in\LL$ and $\tf{g}\in\LLk$ be two normalized wave 
functions such that $f(y)$ and $g(x)$ exist a.e.~and are 
integrable.
If there exist three constants \ul{$\ga >0$}, $C>0$, and 
$L \geq 0$ such that
\be
\abs{f(y)} &\leq& \sur{C}{y^{1+\ga}}\,, 
\saut \mbox{if \ $y \ \leq -L$,}
\label{ScloseNhyp}
\ee 
then for both black holes and for $y_o\gg 1$,
\beaa{rcl}
\left\vert\,\bar{N}_g[f_{y_o}] -\bar{N}_o[f_{y_o}]
\,\right\vert 
&\leq& \no{g}_\LLum{x}^2 \
\sur{4\,C^2}{\ga^2\left(\,\frac{1}{2}\,y_o+L\,\right)^{2\ga}}
\\ [7mm] \hspace{10mm} && +\
e^{2L-y_o}\,\no{f}_\LLu{y}^2 \, \abs{g(0^-)}^2.
\label{ScloseNgNo}
\eeaa
\end{theorem} 
This theorem is proved in appendix \ref{ap:ScloseN}.
Under the specified conditions, it shows that for both 
black-hole models 
\be
\lim_{y_o \rightarrow +\infty} 
\left( \, \bar{N}_g[f_{y_o}] -\bar{N}_o[f_{y_o}] \, \right) 
&=& 0.
\ee
The bound (\ref{ScloseNgNo}) is composed of two terms:
the first decreases algebraically in $y_o$ with exponent 
$2\ga$ and depends only on the restriction of $g$ to the 
{\it exterior} of the black holes;
the second decreases exponentially in $y_o$ and is absent 
if $g$ vanishes at the horizon (more precisely at $x=0^-$).

The corollaries below follow directly from the result 
(\ref{SpNRBH}) and theorems \ref{th:SpCloseN} and 
\ref{th:ScloseN}.
\begin{corollary}
Under the hypothesis of theorem \ref{th:ScloseN}, for the 
relativistic model and for $y_o \gg 1$,
\beaa{rcl}
\left\vert\, 
\bar{N}_g[\fyo{f}]-\bar{N}^{Th}_{2\pi,out}[\fyo{f}] 
\,\right\vert 
&\leq& \no{g}_\LLum{x}^2 \
\sur{4\,C^2}{\ga^2\left(\,\frac{1}{2}\,y_o+L\,\right)^{2\ga}}
\\ [7mm] \hspace{10mm} && +\
e^{2L-y_o}\,\no{f}_\LLu{y}^2 \, \abs{g(0^-)}^2.
\label{corSpCloseN1}
\eeaa
\end{corollary}
\begin{corollary}
Under the hypothesis of theorem \ref{th:ScloseN} and assuming 
further that \ul{$\ga>1/2$} and $L\geq 1$, for the CGHS 
model and for $y_o \gg 1$,
\be
\left\vert\, 
\bar{N}_g[\fyo{f}]-\bar{N}^{Th}_{2\pi,out}[\fyo{f}] 
\,\right\vert 
&\leq& \left(1+\no{g}_\LLum{x}^2 \, \right) \
\sur{64\,C^2}
{\ga \,(2\ga-1)\left(\,\frac{1}{4}\,y_o+L-1\,\right)^{2\ga-1}}
\label{corSpCloseN2} \\ [1mm] \hspace{13mm}
&&+ \ 2\, e^{2L-y_o/2} \, 
\left(\, \no{f}_{\LLu{y}} + \no{f'}_{\LLu{y}}\,\right)^2 \ 
\left(\,1+ \abs{g(0^-)}^2 \,\right).
\nonumber
\ee
\end{corollary}
Under the specified conditions, these corollaries show that 
the mean number of particles stimulated by a one-particle state 
is thermal asymptotically close to the horizon for both 
black-hole models:
\be
\lim_{y_o\rightarrow +\infty}\left(\,
\bar{N}_g[\fyo{f}] - \bar{N}^{Th}_{2\pi,out}[\fyo{f}] 
\,\right) &=& 0.
\label{SgNoNth}
\ee
The specified conditions are stronger for the CGHS model 
than for the relativistic model, i.e. in this last case 
eq.~(\ref{SgNoNth}) is valid for all algebraically decreasing
functions.
{\it For the CGHS model}, theorems \ref{th:SpCloseN} and 
\ref{th:ScloseN} imply that {\it eq.~(\ref{SgNoNth}) is not true
for sufficiently weakly decreasing functions}.
The exponents in the denominator in eqs 
(\ref{corSpCloseN1}) and (\ref{corSpCloseN2}) are $2\ga$ 
for the relativistic model and $2\ga-1$ for the CGHS model.
\subsection{The complex scalar field}
For the complex scalar field, the incoming and outgoing 
two-point functions are also defined respectively by 
eqs (\ref{Sg2PFx}) and (\ref{Sg2PFy}), and
the regularized energy-momentum tensors of the complex scalar 
field in the incoming and outgoing coordinates by eqs 
(\ref{defgTEInormalx}) and (\ref{defgTEInormaly}), or by eqs 
(\ref{defgTEIsubx}) and (\ref{defgTEIsuby}).
\begin{theorem} \label{th:simpleTEIcomplex}
If $\tf{g}\in\ssp$ is a normalized test function, then
for the complex scalar field, between kernels of distributions 
on $\so\times\so$,
\be
W_g (x,x') &=& W_o (x,x')+\sur{1}{4}\,\Re [\,G(x) \, G(x')^*\,],
\label{CS2PFx} \\ [2mm]
\wh{W}_g (y,y') &=& \wh{W}_o (y,y') + 
\sur{1}{4} \, \Re [\,G(x(y)) \, G(x(y'))^*\,],
\label{CS2PFy}
\ee
where $W_o (x,x')$ and $\wh{W}_o (y,y')$ are given by eqs
(\ref{Sp2PFx}) and (\ref{Sp2PFy});
furthermore
\beaa{rcccl}
\mbox{\hspace{4mm}}T_g(x) &=& 
T_{g^*}(x) &=& \sur{1}{4}\, \abs{g(x)}^2, 
\mbox{\hspace{28mm}}
\label{CgTEIx}
\eeaa
\mbox{} \vspace{-12mm} \\
\beaa{rcccl}
\wh{T}_g(y) &=& \wh{T}_{g^*}(y) &=&
\wh{T}_o(y) + x'(y)^2\ T_g(x(y)),
\label{CgTEIy}
\eeaa
where $\wh{T}_o(y)$ is given by eq.~(\ref{SpTEI}).
\end{theorem}
The proof of this theorem is similar to that of theorems
\ref{th:Sg2PF} and \ref{th:simpleTEI}.
Note the appearance of the factor 1/4 in eqs (\ref{CS2PFx}) 
to (\ref{CgTEIx}) instead of 1/2 in eqs 
(\ref{S2PFx}), (\ref{S2PFy}) and (\ref{gTEIx}).
For the complex scalar field, the conclusions this theorem
implies are identical to those for the real scalar field, 
i.e.~theorems \ref{th:2PFBH}, \ref{th:simpleTEIBH} and 
\ref{th:simpleenergies} are also true in the complex case.
The results of subsection \ref{sub:MNCP} concerning the mean 
number of created particles are also valid for the complex 
scalar field.

In the incoming and outgoing coordinates, the mean currents
associated with a test function $g\in\ssp$ are denoted by
$J_{g}(x)$ and $\wh{J}_g(y)$ respectively.
These are given in the normal and subtraction regularization 
schemes by
\be
J_{g} (x) &=& (\gO_g, :\gU(x):_{in} \, \gO_g),
\label{CgcurrentNorx} \\ [2mm]
J_g (x) &=&  \lim_{\gep\rightarrow 0} \, 
\left[\,(\gO_g,\gU_\gep (x)\,\gO_g)
-(\vac,\gU_\gep (x)\,\vac)\,\right],
\label{CgcurrentSubx} \\ [2mm]
\wh{J}_g(y) &=& (\gO_g, :\wh{\gU}(y):_{out} \, \gO_g),
\label{CgcurrentNory} \\ [2mm]
\wh{J}_g(y) &=& \lim_{\gep\rightarrow 0} \, 
\left[ \, (\gO_g, \wh{\gU}_\gep (y) \, \gO_g)
- (\vac,\gU_\gep (x(y)) \, \vac) \, \right],
\label{CgcurrentSuby}
\ee
where $\gU(x)$, $\wh{\gU}(y)$, $\gU_\gep (x)$ and
$\wh{\gU}_\gep (y)$ are defined by eqs (\ref{currentx}) to 
(\ref{currentybis}).
The incoming and outgoing total charges of the state $\gO_g$ 
are given by the integrals
\beaa{rcccccl}
Q^{in}_g &=&  \intii dx \, J_g (x),
&\saut&
Q^{out}_g &=&  \intii dy \ \wh{J}_g (y).
\label{CgCharge}
\eeaa 
\begin{theorem} \label{th:CgCurrentx}
If $\tf{g}\in\ssp$ is a normalized test function, and if
$J_{g}(x)$ and $\wh{J}_g(y)$ are defined by eqs
(\ref{CgcurrentNorx}) and (\ref{CgcurrentNory}), or
by eqs (\ref{CgcurrentSubx}) and (\ref{CgcurrentSuby}), then
\beaa{rcccl}
J_g (x) &=& - J_{g^*} (x) &=&
- \sur{\I}{4} \, G(x)^*  
\stackrel{\leftrightarrow}{\partial}_x G(x), 
\label{CgCurrentx}
\eeaa
\mbox{} \vspace{-11mm} \\
\beaa{rcccl}
\mbox{\hspace{11mm}}\wh{J}_g (y) &=& - \wh{J}_{g^*} (y) 
&=& - \sur{\I}{4} \, 
G(x(y))^*  \stackrel{\leftrightarrow}{\partial}_y G(x(y)),
\label{CgCurrenty} 
\eeaa
where $G$ is the primitive (\ref{defG}) of g;
furthermore
\beaa{rcccl}
Q^{in}_g &=& - Q^{in}_{g^*} &=& 1,
\label{CgChargex}
\eeaa
and for both models,
\beaa{rcccl}
Q^{out}_g &=& - Q^{out}_{g^*} &=& \intio dx \, J_g (x).
\label{CgChargey} 
\eeaa
\end{theorem}
This theorem is proved in appendix \ref{ap:CgCurrentx}.
It shows that the mean current is a non-local function, and
that the mean current of a particle is not positive definite 
locally, nor is that of an antiparticle negative definite 
locally\footnote{
In contrast to the case of the Dirac field, 
see ref.~\cite{GaW}.}.
For both models, the outgoing mean charge (\ref{CgChargey}) 
is only part of the incoming mean charge (\ref{CgChargex}).
If $Q^{in}$ is positive, $Q^{out}$ may be negative,
so {\it it is possible to observe a negative outgoing mean 
charge for a positive incoming mean charge}.
\section{Emission stimulated by a thermal state}
In this section we assume that the incoming state is thermal
with temperature $\gb^{-1}$.
This temperature must not be confused with that of the emitted 
radiation.
The mean value of an observable $A$ in this thermal state
is given by
\be
\mtin{A} &=& \sur{\Trin \left[\, e^{-\gb H_{in}} \, A \,\right]}{
\Trin \left[\, e^{-\gb H_{in}}\,\right]}
\ee
and is thus a thermal mean value in the incoming Hilbert space $\hin$.
The definitions and results of this section are valid for both
real and complex scalar fields.
\subsection{The two-point function}
The incoming and outgoing two-point functions for an incoming thermal 
state, denoted by $W_{\gb,in}^{Th}(x,x')$ and 
$\wh{W}_{\gb,in}^{Th}(y,y')$ respectively, are defined as
\be
W_{\gb,in}^{Th}(x,x') &=&
\mtin{\phi(x)\, \phi(x')^\dagger}\,, \\ [2mm]
\wh{W}_{\gb,in}^{Th}(y,y') &=&
\mtin{\ch{\phi}(y)\, \ch{\phi}(y')^\dagger}\,,
\ee
and are related by
\be
\wh{W}_{\gb,in}^{Th}(y,y') &=& W_{\gb,in}^{Th}\,(x(y),x(y')).
\ee
\begin{theorem}
Between kernels of distributions on $\so\times\so$,
\be
W_{\gb,in}^{Th}\, (x,x') &=&
- \sur{1}{4\pi} \log \left\{\, \sur{\gb}{\pi} \,
\sinh \left[\,\frac{\pi}{\gb}\,\left(x'-x+\I\,0^+\right) 
\,\right]\,\right\}, 
\label{Th2PFx}\\ [2mm]
\wh{W}_{\gb,in}^{Th}(y,y')
&=& -\frac{1}{4\pi} \log \left\{ \, \frac{\gb}{\pi} \, 
\sinh \left[\,\frac{\pi}{\gb}\,\left(\,x(y')-x(y)+\I\,0^+\,\right)
\,\right] \right\}.
\label{Th2PFy}
\ee
\end{theorem}
\Proof
See eq.~(\ref{2PFthermal}).
\hfill{$\Box$} \\ \\
This last theorem applies to both black-hole models.
\begin{theorem} \label{th:Th2PF.BH}
Between kernels of distributions on $\so\times\so$,
for both models and for all $\gb > 0$,
\be
\wh{W}_{\gb,in}^{Th}(y,y') &\approx& W_{2\pi,out}^{Th}(y,y'),
\saut \mbox{if \ $y,y'\gg 1$};
\label{Th2PF.BHA} 
\ee
furthermore for the CGHS model and for all $\gb > 0$,
\be
\wh{W}_{\gb,in}^{Th}(y,y') &\approx& W_{\gb,out}^{Th}(y,y'), 
\saut \hspace{2mm}\mbox{if \ $-y,-y'\gg 1$},
\label{Th2PF.BHB}
\ee
where $W_{\gb,out}^{Th}(y,y')$ is given by eq.~(\ref{2PFthermal}).
\end{theorem}
\Proof
For both models, we have
$x(y)\approx -e^{-y}\approx 0^-$ if $ y \gg 1$.
Together with eq.~(\ref{Sp2PFy}), eq.~(\ref{Th2PFy}) then implies 
for all $\gb>0$, if $y,y'\gg 1$,
\beaa{rcccl}
\wh{W}_{\gb,in}^{Th}(y,y')
& \approx &-\sur{1}{4\pi} \log 
\left(\,e^{-y}-e^{-y'}+\I\,0^+\,\right)
&\approx& \wh{W}_o(y,y'),
\label{eq:KMSAstimule}
\eeaa
from which eq.~(\ref{Th2PF.BHA}) is deduced using the results
(\ref{Sp2PFRBH}) and (\ref{Sp2PFCGHSa}).

For the CGHS model, we have further that $x(y)\approx y$ 
if $-y\gg 1$.
Equation (\ref{Th2PFy}) then implies for all $\gb>0$, if 
$-y,-y'\gg 1$,
\beaa{rcl}
\wh{W}_{\gb,in}^{Th}(y,y') &\approx& 
-\sur{1}{4\pi} \log \left\{ \, \sur{\gb}{\pi} \, \sinh 
\left[\,\sur{\pi}{\gb}\,\left(y'-y+\I\,0^+\right)\,\right] \right\} 
\label{eq:KMSBstimule}
\eeaa
from which eq.~(\ref{Th2PF.BHB}) is deduced using 
eq.~(\ref{2PFthermal}).
\hfill{$\Box$} \\ \\
For both black-hole models and in the outgoing coordinates, 
the incoming thermal state is thus thermal close to the horizon.
In this region, the associated outgoing temperature {\it does not 
depend} on the incoming temperature and coincides with that 
for the spontaneous emission in the same region.
Furthermore, for the CGHS model only, and in the outgoing
coordinates, the incoming thermal state of temperature $\gb^{-1}$ 
is also thermal very far from the horizon, and the associated outgoing
temperature is also $\gb^{-1}$.
\subsection{The energy-momentum tensor}
In the incoming and outgoing coordinates, the energy-momentum tensors 
for an incoming thermal state of temperature $\gb^{-1}$ will be 
denoted by $T^{Th}_{\gb,in}(x)$ and $\wh{T}_{\gb,in}^{Th}(y)$ 
respectively.
These are given in the normal and subtraction regularization schemes
by
\be
T_{\gb,in}^{Th}(x) &=& \mtin{ \,:\gT(x):_{out}\,}\,, 
\label{ThdefTEInorx}\\[2mm]
T_{\gb,in}^{Th}(x) &=& \lim_{\gep\rightarrow 0} \, 
\left[ \, \mtin{ \gT_\gep(x)} 
- (\vac,\gT_\gep (x) \, \vac) \, \right],
\label{ThdefTEIsubx} \\ [2mm]
\wh{T}_{\gb,in}^{Th}(y) &=& \mtin{ \,:\wh{\gT}(y):_{out}\,}\,, 
\label{ThdefTEInory}\\[2mm]
\wh{T}_{\gb,in}^{Th}(y) &=& \lim_{\gep\rightarrow 0} \, 
\left[ \, \mtin{ \wh{\gT}_\gep (y)} 
- (\vac,\gT_\gep (x(y)) \, \vac) \, \right],
\label{ThdefTEIsuby}
\ee
where the observables $\gT(x)$, $\gT_\gep (x)$, $\wh{\gT}(y)$ and 
$\wh{\gT}_\gep (y)$ are defined by eqs (\ref{TEIopx}) to 
(\ref{TEIybis}).
\begin{theorem} \label{th:ThTEI}
If $T_{\gb,in}^{Th}(x)$ and $\wh{T}_{\gb,in}^{Th}(y)$ are defined by 
eqs (\ref{ThdefTEInorx}) and (\ref{ThdefTEInory}), or by eqs
(\ref{ThdefTEIsubx}) or (\ref{ThdefTEIsuby}), then
\be
T^{Th}_{\gb,in}(x) &=& T^{Th}_{\gb,in} \ = \ 
\frac{\pi}{12 \gb^2}\,, 
\label{ThTEIx} \\ [2mm]
\wh{T}_{\gb,in}^{Th}(y) &=& 
\wh{T}_o(y)+ x'(y)^2 \ T^{Th}_{\gb,in}\,,
\label{ThTEIy}
\ee
where $\wh{T}_o(y)$ is given by eq.~(\ref{SpTEI}).
\end{theorem}
This last theorem is proved in appendix \ref{ap:ThTEI}.
Theorem \ref{th:simpleenergies}, which concerns the incoming 
and outgoing energies, may be generalized to the incoming thermal 
states.
Theorem \ref{th:ThTEI} applies to both black-hole models.
\begin{theorem}
For both models,
\be
\lim_{y\rightarrow +\infty} 
\wh{T}_{\gb,in}^{Th}(y) &=& T_{2\pi,out}^{Th}\,,
\label{ThBhTEIa}
\ee
furthermore, for the CGHS model,
\be
\lim_{y\rightarrow -\infty} 
\wh{T}_{\gb,in}^{Th}(y) &=&  T_{\gb,out}^{Th}\,,
\label{ThBhTEIb}
\ee
where $T_{\gb,out}^{Th}$ is given by eq.~(\ref{thermalTEI}).
\end{theorem}
\Proof
For both models, $x'(y)\approx 0$ if $y\gg 1$.
For the CGHS model, it is also true that $x'(y)\approx 1$ if 
$-y\gg1$.
Equations (\ref{ThBhTEIa}) and (\ref{ThBhTEIb}) are then deduced
from eq.~(\ref{ThTEIy}) using the results (\ref{SpTEIRBH}), 
(\ref{SpTEICGHSa}) and (\ref{SpTEICGHSb}).
\hfill{$\Box$} \\ \\
For both black-hole models, the behavior of the energy-momentum
tensor is thus thermal close to the horizon.
In this region, the outgoing temperature is {\it not affected}
by the incoming temperature and coincides with that for
the spontaneous emission in the same region.
Furthermore, for the CGHS model only, the energy-momentum
tensor is also thermal very far from the black hole and its associated
temperature is equal to the temperature of the incoming thermal state.
\subsection{The mean number of created particles}
The mean number $\bar{N}_{\gb,in}^{Th}[f]$ of particles
induced by a thermal state of temperature $\gb^{-1}$ is defined as
\be 
\bar{N}_{\gb,in}^{Th}[f] &=& 
\mtin{\ch{\phi}[f]^\dagger\,\ch{\phi}[f]}\,.
\label{ThNdef}
\ee
\begin{theorem} \label{th:ThN}
If $\tf{f}\in\LL$ is a normalized wave function such that $f(y)$
exists a.e.~and is integrable, then
\be
\bar{N}_{\gb,in}^{Th}[f] &=&
\intii \sur{dk}{2k}\,\sur{\abs{(U \tf{f})(k)}^2}{e^{\gb k}-1}\,,
\label{ThNd}
\ee
and for all $\gb>0$,
\be
\bar{N}_o[f]  &\leq& \bar{N}_{\gb,in}^{Th}[f].
\label{ThNb}
\ee
\end{theorem}
\Proof
Equation (\ref{ThNd}) is deduced directly from def.~(\ref{ThNdef})
(see ref.~\cite{FV} for details), and implies, with 
eq.~(\ref{number}),
\be 
\bar{N}_{\gb,in}^{Th}[f] &=& \bar{N}_o[f] 
+\intoi \sur{dk}{2k} \, \sur{1}{e^{\gb k}-1} \,
\left[ \, \abs{(U\tf{f})(k)}^2 +\abs{(U\tf{f})(-k)}^2\, \right],
\label{ThNa}
\ee
from which eq.~(\ref{ThNb}) is deduced.
\hfill{$\Box$} \\ \\
The mean number of particles stimulated by a thermal state is thus
always equal to or larger than that for the spontaneous emission.
Equation (\ref{ThNa}) implies that
\be
\lim_{\gb\rightarrow\infty} \bar{N}_{\gb,in}^{Th}[f]
&=& \bar{N}_o[f].
\ee
In the incoming momentum, the integral in eq.~(\ref{ThNd}) may be 
infrared divergent.
For the relativistic black-hole model, it is always divergent.
\begin{theorem} \label{th:ThRBhN}
If $\tf{f}\in\LL$ is a normalized wave function such that $f(y)$
exists a.e.~and is integrable, then for the relativistic model
\be
\bar{N}_{\gb,in}^{Th}[f]- \bar{N}_o [f] &=& \infty,
\saut \forall\,\gb>0,
\label{ThRBhN}
\ee
and so $\bar{N}_{\gb,in}^{Th}[f]$ is always infinite for all $\gb>0$.
\end{theorem}
\Proof
For the relativistic black hole, the kernel of $U$ 
is given by eq.~(\ref{UkernelTNR}).
Using eq.~(\ref{ThNd}), the divergent term,
\be
\intoi \sur{dk}{2k} \ 
\sur{e^{\I\,\log k \,(p-p')}}{e^{\gb k}-1}
&=& \infty, \saut \forall \, p,p'\in \R_+,
\ee
appears in the kernel of $\bar{N}_{\gb,in}^{Th}[f]$.
This implies eq.~(\ref{ThRBhN}).
\hfill{$\Box$} \\ \\
For the relativistic black-hole model and for all inverse temperature, 
this theorem also shows that the incoming thermal state induces an 
infinite number of extra particles in comparaison with the 
spontaneous case and for a given test function.
If the mean number of stimulated particles is considered for a given 
mode or in total (see def.~(\ref{Ntot})), this last result is true 
for {\it all models}.
\begin{theorem} \label{th:ThmodeN}
If the outgoing function $f_p$ is defined by
eq.~(\ref{Sdefmode}), then for any transformation of coordinates 
$x=x(y)$ and for $p\geq0$
\be
\bar{N}_{\gb,in}^{Th}[f_p] - \bar{N}_o[f_p] &=& \infty, 
\label{ThmodeN} \\ [2mm]
\bar{N}_{\gb,in}^{Th,tot} - \bar{N}_o^{tot} &=& \infty.
\label{ThtotN}
\ee
\end{theorem}
\Proof
It is alway true that
\be
\ind{\tfch{f}}{p}(k) &=& 2p\,U(k,p).
\ee
Using eq.~(\ref{ThNd}), the divergent term
\be
\intii \sur{dk}{2k}\,
\sur{e^{\I\,k\left[\,x(y)-x(y')\,\right]}}{e^{\gb k}-1}
&=& \infty, \saut \forall\,y,y'\in\R,
\ee
appears in the kernel of $\bar{N}_{\gb,in}^{Th}[f_p]$ for all 
transformations $x=x(y)$.
This implies eqs (\ref{ThmodeN}) and (\ref{ThtotN}).
\hfill{$\Box$}

For the CGHS black-hole model, bounds for 
\mbox{$\bar{N}_{\gb,in}^{Th}[f] - \bar{N}_o[f]$} and  
$\bar{N}_{\gb,in}^{Th}[f]$ exist in terms of the Fourier
transform $\tf{f}$.
\begin{theorem} \label{th:Thbound}
If $\tf{f}\in\LL$ is a normalized wave function such that $f(y)$
exists a.e.~and is integrable, then for the CGHS model
\be
\left\vert\,\bar{N}_{\gb,in}^{Th}[f] - \bar{N}_o[f]\,\right\vert 
&\leq& \sur{\pi}{\gb} \, 
\intoi \sur{dp}{2p} \, \sur{\abs{\tf{f}(p)}^2 }{1-e^{-2\pi p}}\cdot
\label{Thbound}
\ee
\end{theorem}
This theorem is proved in appendix \ref{ap:Thbound} and, together
with theorem \ref{th:SpBoundN}, it implies the corollary below.
\begin{corollary}
If $\tf{f}\in\LL$ is a normalized wave function such that $f(y)$
exists a.e.~and is integrable, then for the CGHS model
\be
\bar{N}_{\gb,in}^{Th}[f] &\leq& C \, (1+\gb^{-1}) \, 
\intoi \sur{dp}{2p} \, \sur{\abs{\tf{f}(p)}^2 }{1-e^{-2\pi p}}\,,
\ee
where $C>0$ is a constant.
\end{corollary}
\subsection{Close to the horizon}
For the relativistic black-hole model and if $f_{y_o}$ is defined
by eq.~(\ref{deffyo}), it is clear that
$\bar{N}_{\gb,in}^{Th}[f_{y_o}]$ does not approach
$\bar{N}_o[f_{y_o}]$ asymptotically near the horizon, since 
$\bar{N}_{\gb,in}^{Th}[f_{y_o}]$ is alway infinite for this model
(see theorem~\ref{th:ThRBhN}).
However, for the CGHS black-hole model and under some conditions, 
$\bar{N}_{\gb,in}^{Th}[f_{y_o}]$ does indeed tend to
$\bar{N}_o[f_{y_o}]$ asymptotically near the horizon:
\be 
\lim_{y_o \rightarrow +\infty} 
\left( \ \bar{N}_{\gb,in}^{Th}[\fyo{f}]-\bar{N}_o[\fyo{f}]
\ \right) &=& 0.
\label{ThNthNo}
\ee
This is shown in the following theorem, which is proved in appendix 
\ref{ap:ThCGHSNhorizon}.
\begin{theorem} \label{th:ThCGHSNhorizon}
Let $\tf{f}\in\LL$ be a normalized wave function such that $f(y)$
exists a.e.~and is integrable.
If there exist three constants \ul{$\ga > 1/2$}, $C>0$ and 
$L \geq 1$ such that
\be
\abs{f(y)} &\leq& \sur{C}{y^{1+\ga}}\,, 
\saut \mbox{if \ $y \ \leq -L$,}
\label{HYPThCGHSNhorizon}
\ee 
then, for the CGHS model and for $y_o > 0$,
\beaa{rcl}
\left\vert\, \bar{N}_{\gb,in}^{Th}[f_{y_o}] - \bar{N}_o[f_{y_o}] 
\,\right\vert 
&\leq& \sur{C^2}
{\gb\,\ga^2 \,(2\ga-1)\left(\frac{1}{2}\,y_o+L\right)^{2\ga-1}} 
\\ [7mm] &&
+ \ \sur{4\pi}{\gb}\ e^{L-y_o/2}\, \intoi \sur{dp}{2p}\,
\sur{\abs{\tf{f}(p)}^2 }{1-e^{-2\pi p}}\cdot
\label{ThCGHSNhorizon}
\eeaa
\end{theorem} 

For the relativistic black-hole model, theorem \ref{th:ThRBhN}
implies that
\be 
\bar{N}_{\gb,in}^{Th}[\fyo{f}]-
\bar{N}_{2\pi,out}^{Th}[\fyo{f}]  &=& \infty,
\saut \forall\,y_o\in\R.
\ee
{\it In this case the mean number of created particles is thus never 
thermal close to the horizon}.
For the CGHS black-hole model and under some conditions,
the following corollary to theorems \ref{th:SpCloseN} and 
\ref{th:ThCGHSNhorizon} shows that $\bar{N}_{\gb,in}^{Th}[f]$ 
tends to the thermal value $\bar{N}^{Th}_{2\pi,out}[f]$ close to the 
horizon:
\be 
\lim_{y_o \rightarrow +\infty} 
\left( \ \bar{N}_{\gb,in}^{Th}[\fyo{f}]-
\bar{N}_{2\pi,out}^{Th}[\fyo{f}] \ \right) = 0.
\label{ThNthNth}
\ee
\begin{corollary} \label{th:ThCGHSNhorizonBis}
Under the hypothesis of theorem \ref{th:ThCGHSNhorizon},
for the CGHS model and for $y_o >0$,
\be
\left\vert \, \bar{N}_{\gb,in}^{Th}[\fyo{f}] -
\bar{N}^{Th}_{2\pi,out}[\fyo{f}] \, \right\vert &\leq&
\sur{32\,(1+\gb^{-1})\,C^2}
{\ga^2(2\ga-1)\left(\frac{1}{4}\,y_o+L-1\right)^{2\ga-1}}
\hspace{50mm}
\label{ThCGHhorizonBis}
\ee
\mbox{} \vspace{-9mm} \\
\be
\hspace{45mm}
&& + \ e^{2L-y_o/2} \left[\, \left( \, \Lu{f}+ \Lu{f'} \right)^2  
+ \sur{4\pi}{\gb}
\intoi \sur{dp}{2p} \,\sur{\abs{\tf{f}(p)}^2 }{1-e^{-2\pi p}}\,
\right].
\nonumber
\ee
\end{corollary}
{\it For the CGHS black-hole model}, theorem \ref{th:SpCloseN} 
implies that {\it eqs (\ref{ThNthNo}) and (\ref{ThNthNth}) are not 
valid if $f$ decreases sufficiently weakly very far from the 
horizon and does not oscillate}.
\section{Conclusions}
The emission of massless bosons by the relativistic and CGHS 
black holes have been studied for one-particle and thermal 
incoming states.
Mean values of observables constructed in the outgoing 
coordinates were computed in these states.
The results obtained in this paper are summarized in table 1.
They show that the emitted radiation exhibits both thermal and 
non-thermal properties close to the horizon for both black-hole 
models.
For all incoming states, the thermal properties are always 
associated with the temperature $\frac{M}{2\pi}$ for the 
relativistic black hole and with the temperature 
$\frac{\gl}{2\pi}$ for the CGHS black hole.
\begin{figure}
\begin{center}
\begin{tabular}{|c||c|c|c|}
\hline 
&&& \\ [-2mm]
Relativistic BH & $\wh{W}(y,y')$ & $\wh{T}(y)$ & $\bar{N}[f]$ \\ 
&&& \\ [-2mm]
\hline 
\mbox{\hspace{33mm}}& \mbox{\hspace{20mm}}&
\mbox{\hspace{20mm}}&\mbox{\hspace{20mm}} \\ [-4.6mm] \hline
\rule[-5.5mm]{0mm}{12mm} 
vacuum & \oo$^1$ & \oo$^1$ & \oo$^1$ \\
\hline
\rule[-5.5mm]{0mm}{12mm} 
one-particle state & $\otimes$ & \oo & \oo$^2$ \\
\hline
\rule[-5.5mm]{0mm}{12mm} 
thermal state & \oo & \oo & $\times$  \\
\hline
\end{tabular}
\\ \rule{0mm}{10mm} \\
\begin{tabular}{|c||c|c|c|}
\hline 
&&& \\ [-2mm]
CGHS BH & $\wh{W}(y,y')$ & $\wh{T}(y)$ & $\bar{N}[f]$ \\
&&& \\ [-2mm]
\hline 
\mbox{\hspace{33mm}}& \mbox{\hspace{20mm}}&
\mbox{\hspace{20mm}}&\mbox{\hspace{20mm}} \\ [-4.6mm] \hline
\rule[-5.5mm]{0mm}{12mm} 
vacuum & \oo & \oo & $\otimes$ \\
\hline
\rule[-5.5mm]{0mm}{12mm} 
one-particle state & $\otimes$ & \oo & 
$\otimes$ \\
\hline
\rule[-5.5mm]{0mm}{12mm} 
thermal state & \oo & \oo & $\otimes$  \\
\hline
\end{tabular} 
%
%
\\ \rule{0mm}{10mm} \\ 
\parbox[t]{130mm}{Table 1: Behavior of mean values of the 
outgoing observables in a given incoming state, close to the 
horizon of the relativistic and CGHS black holes. \\
\oo: thermal; $\times$: non-thermal;
$\otimes$: thermal or non-thermal depending on the test functions;
1: everywhere, i.e.~not only near the horizon;
2: in general.} 
\end{center}
\end{figure}

For the {\it relativistic} black hole and for non-vacuum 
incoming states, the emitted radiation has no thermal 
properties, except close to the horizon, contrary to the 
spontaneous emission.
For this model, the mean number of created particles stimulated 
by a thermal state is {\it non-thermal} for all outgoing wave
functions.
For the {\it CGHS} black-hole model and for all the 
incoming states studied, the mean number of particles may be 
non-thermal for some outgoing test functions.

For {\it one-particle} incoming states and for both black-hole 
models, the outgoing two-point function may be thermal or not, 
depending on the incoming test function.
For incoming {\it thermal} states and for both models, 
the outgoing two-point function is thermal close to the horizon 
for all incoming temperatures.

For both black-hole models, it is remarkable that the 
{\it energy-momentum tensor} is thermal close to the horizon for 
all the incoming states considered.
It thus seems to be stable with respect to the incoming state.
In particular, the energy of the incoming state must be very 
large near the horizon to modify it close to the horizon.
For both black-hole models, a non-vacuum incoming state amplifies 
the emitted radiation,
in the sense that the energy-momentum tensor and mean number
of particles are equal to or larger than those for the 
spontaneous emission.
\\ \\
{\large \bf Acknowledgments} 
\\ \\
I thank G.~Wanders for stimulating discussion and criticism,
and D.~Rickebusch for correcting the manuscript.
\appendix\section{Appendices}
\begin{theorem} \label{th:analyticity}
If $\tf{g} \in L^q (dk,\R_+) $ with $q \in [1,2] $ satisfies 
\be
\tf{g}(k) &=& \gt(k) \, \tf{g}(k), \saut \forall\,k\in\R,
\ee
then
\begin{enumerate}
\item[i)]  $g$ is analytic in the upper complex half-plane
$\Im(x)>0$,
\item[ii)] there exists a positive constant $C$ such that
\be
\abs{g(x)}\ < C, \saut \mbox{ if \ $\Im(x) > 0$,} 
\ee
\end{enumerate}
and so $g$ is regular in the upper complex half-plane
$\Im(x)>0$.
\end{theorem}
This theorem is proved in ref.~\cite{Ch}.
The primitives $\cha{F}(x)$ and $F(y)$ of $\ch{f}(x)$ and $f(y)$
are defined respectively as
\be
\cha{F}(x) &=& \int_{x(-\infty)}^x dx'\, \ch{f}(x'),
\label{eq:defprimitivex} \\ [2mm]
F(y) &=& \int_{-\infty}^y dy'\, f(y').
\label{eq:defprimitivey}
\ee
\sub{Proof of theorem \ref{th:simpleTEI}} 
\label{ap:simpleTEI}
Assuming that $T_g(x)$ and $\wh{T}_g(y)$ are defined 
by eqs (\ref{defgTEInormalx}) and (\ref{defgTEInormaly}) 
respectively,
the incoming and outgoing fields are expanded as
\be
\phi(x) &=& \sur{1}{\sqrt{2\pi}} \,
\intoi \sur{dk}{2k} \, \left[\, 
a_{in}(k)\,e^{-\I kx} + a_{in}^\dagger(k)\,e^{\I kx}\,\right],
\label{devphiin} \\ [2mm]
\ch{\phi}(y) &=& \sur{1}{\sqrt{2\pi}} \,
\intoi \sur{dp}{2p} \, \left[\, 
a_{out}(p) \, e^{-\I py} + a_{out}^\dagger(p)\,e^{\I py}\,\right].
\label{devphichapout}
\ee
Ordering normally the field operators in 
def.~(\ref{defgTEInormaly}), using
\be
(\vac, a_{out}(p) \, a_{in}^\dagger(k) \, \vac) &= & 2p\,U(k,p),
\\ [2mm]
(\vac, a_{out}^\dagger(p) \, a_{in}^\dagger(k) \, \vac) &=& 
2p\,U(k,-p),
\ee
and
\be
\intoi dp\, 2p\,U(k,p)\ e^{-\I yp} &=& 
-\sur{1}{\pi} \intii dy'\,
\sur{e^{-\I kx(y')}}{\left(\,y-y'-\I\,0^+ \,\right)^2}\,,
\ee
it follows that
\be
\wh{T}_g(y) &=& (\vac, :\wh{\gT}(y):_{out} \, \vac)
\ + \ \sur{x'(y)^2}{8\pi^2} \,\left\vert \, 
\intii dx \, \sur{g(x)}{x-x(y)-\I\,0^+} \,\right\vert^2\cdot
\label{eq:simpleTEIa}
\ee
Using the Cauchy theorem and def.~(\ref{SpTEI}), this last 
equation implies that
\be
\wh{T}_g(y) &=& \wh{T}_o(y) \ 
+ \ \sur{x'(y)^2}{2} \, \abs{g(x(y))}^2,
\label{eq:simpleTEIb}
\ee
since $g$ is a regular function in the upper complex half-plane
$\Im(x)>0$ (see theorem \ref{th:analyticity}).
Equations (\ref{gTEIx}) and (\ref{gTEIy}) then follow from 
eq.~(\ref{eq:simpleTEIb}) with $x(y)=y$.

Alternatively, if $T_g(x)$ and $\wh{T}_g(y)$ are defined 
by eqs (\ref{defgTEIsubx}) and (\ref{defgTEIsuby}),
eqs (\ref{gTEIx}) and (\ref{gTEIy}) are directly obtained
using def.~(\ref{SpTEI}), Wick's theorem and
\be
(\vac, \phi[g]\,\partial_y\ch{\phi}(y)\,\vac) &=& 
\sur{\I}{2}\,\partial_y G(x(y)).
\ee
\hfill{$\Box$}
\sub{Proof of theorem \ref{th:Sbound}}
\label{ap:Sbound}
Considering the term $\bra\tf{g},A \tf{f}\ket$ in eq.~(\ref{SN}),
we have
\beaa{rcccl}
\intoi \sur{dk}{2k} \, \tf{g}(k)^* \, \tfch{f}(k) 
&=&  \sur{\I}{2}\intii dk \, \tf{g}(k)^* \, \tfcha{F}(k) 
&=&  \sur{\I}{2}\intio dx\, g(x)^* \, \cha{F}(x) \\ [4mm]
&=&  \sur{\I}{2}\intii dy \, \hc{g}(y)^* \, F(y) 
&=& \intoi \sur{dp}{2p} \, \tfhc{g}(p)^* \, \tf{f}(p),
\label{SboundA}
\eeaa
where $\cha{F}(x)$ and $\hc{g}(y)$ are defined by 
eqs (\ref{eq:defprimitivex}) and (\ref{Sdefgoutgoing}) 
respectively.
Note that $\tfhc{g}(0)$ may not be zero.
A bound for $\tfhc{g}(p)$ is given by
\beaa{rcccl}
\sqrt{2\pi}\,\abs{\tfhc{g}(p)} &\leq& \no{\hc{g}}_{\LLu{y}}
&\leq& \no{g}_{\LLum{x}}\,.
\label{SboundB}
\eeaa
Equation (\ref{SboundA}) then shows that
\be
\sqrt{2\pi}\,\abs{\bra\tf{g},A \tf{f}\ket} &\leq &
\no{g}_{\LLum{x}} \, \intoi \sur{dp}{2p} \, \abs{\tf{f}(p)}.
\label{SboundC}
\ee
Since the bound (\ref{SboundC}) is also valid for 
$\bra\tf{g}^*,B \tf{f}\ket$, eq.~(\ref{Sbound}) is true.
\hfill{$\Box$}
\sub{Proof of theorem \ref{th:SNmode}}
\label{ap:SNmode} 
Using def.~(\ref{Sdefmode}), eqs (\ref{SN}) and 
(\ref{SboundA}) imply that
\be
\bar{N}_g[f_p] &=& \bar{N}_o[f_p] + 2\,\abs{\tfhc{g}(p)}^2.
\label{ap:SNmodeA} 
\ee
This last equation implies eqs (\ref{SNmode}) and (\ref{SNtotA})
using the bound (\ref{SboundB}) and def.~(\ref{Ntot}).

As $\sqrt{2\pi}\,\tfhc{g}(0)=G(0)$, it is clear from 
eq.~(\ref{SNtotA}) that $\bar{N}_g^{tot} - \bar{N}_o^{tot}$ 
diverges if $G(0)\not=0$.
If $G(0)=0$, integrating by parts gives
\be
\sqrt{2\pi}\,\tfhc{g}(p) &=&
\I\,p \,\intio dx \,G(x)\,y'(x)\,e^{-\I py(x)}.
\ee
For $g\in\ssp$, since $G(x) = \OO(x)$ if $x\approx 0^-$ and 
$G(x)= \OO(1/x)$ if $-x\gg1$, we conclude that 
$\tfhc{g}(p) = \OO(p)$ if $p\approx 0$ (since $y'(x)=\OO(1/x)$ 
if $x\approx 0^-$ and $-x\gg1$ for both black-hole models).
Furthermore $\tfhc{g}(p)=\OO(1/p)$ at infinity. 
This shows that the norm $\no{\tfhc{g}}$ is finite if $G(0)=0$
and $g\in\ssp$.
\hfill{$\Box$}
\sub{Proof of theorem \ref{th:ScloseN}}
\label{ap:ScloseN}
Let $\xi$ be a function such that
\beaa{rcccl}
\Lim_{y_o \rightarrow +\infty} \xi(y_o) &=& 
\Lim_{y_o \rightarrow +\infty} \, [\,y_o-\xi(y_o)\,] 
&=& +\infty.
\eeaa
Considering the term $\bra\tf{g},A \fy{\tf{f}}\ket$ in 
eq.~(\ref{SN}),
we have
\beaa{lll}
\mbox{\hspace{-10mm}}
\left\vert \, \intoi \sur{dk}{k} \, \tf{g}(k)^* \, 
\fy{\tfch{f}}(k) \, \right\vert 
&=&\left\vert \,\intii dy \, 
x'(y)\, g(x(y))^* \, F(y-y_o) \,\right\vert
\mbox{\hspace{35mm}} 
\eeaa
\be
&\leq&
\left\vert \, \Int_{-\infty}^{-\xi(y_o)} 
dy \, x'(y+y_o) \, g(x(y+y_o))^* \, F(y) \, \right\vert +  
\left\vert \, \Int^{+\infty}_{-\xi(y_o)} 
dy \, x'(y+y_o) \, g(x(y+y_o))^* \, F(y) \, \right\vert.
\nonumber
\ee
The two terms in this last equation are treated separately.
On one hand, the hypothesis (\ref{ScloseNhyp}) implies, after an 
integration by parts,
\be
\mbox{\hspace{-10mm}}
\left\vert \, \Int_{-\infty}^{-\xi(y_o)} 
dy \, x'(y+y_o) \, g(x(y+y_o))^* \, F(y) \,\right\vert & &
\mbox{\hspace{70mm}} 
\nonumber
\ee
\mbox{}\vspace{-11mm} \\
\be
&\leq& \, \abs{G(x(y_o-\xi(y_o)))} \, \abs{F(-\xi(y_o))} 
+ \, \no{g}_\LLum{x} \, \Int_{-\infty}^{-\xi(y_o)} dy \, \abs{f(y)} 
\nonumber \\ [1mm]
&\leq& 2\, \no{g}_\LLum{x} \ \sur{C}{\ga \, \xi(y_o)^\ga}\,,
\saut \mbox{if \ $\xi(y_o)\geq L$,}
\label{eq:ScloseNa}
\ee
and on the other hand,
\be
\mbox{\hspace{-7mm}}
\left\vert \,\Int^{+\infty}_{-\xi(y_o)} 
dy \, x'(y+y_o) \, g(x(y+y_o))^* \, F(y) \,\right\vert &\leq& 
\no{f}_\LLu{y} \  
\Int_{x(y_o-\xi(y_o))}^0 dx \, \abs{g(x)}.
\mbox{\hspace{10mm}}
\label{eq:ScloseNb}
\ee
An identical bound is obtained for the term 
$\bra\tf{g}^*,B \fy{\tf{f}}\ket$.

With $\xi(y_o)=\sur{1}{2}\,y_o+L$, eqs (\ref{eq:ScloseNa}) and 
(\ref{eq:ScloseNb}) imply if $y_o \geq 0$
\beaa{rcl}
\left\vert\,\bar{N}_g[f_{y_o}] -\bar{N}_g[f_{y_o}]
\,\right\vert &\leq&
4\, \no{g}_\LLup{x}^2 \,
\sur{C^2}{\ga^2\left(\,\sur{1}{2}\,y_o+L\,\right)^{2\ga}} 
\\ [7mm]
&& +\ \no{f}_\LLu{y}^2 \, x(L-y_o/2)^2 \, M_g (x(L-y_o/2))^2,
\label{eq:ScloseNc}
\eeaa
where $M_g(x)$ $(x< 0)$ is the average of $\abs{g}$ on the
interval $[\,x,\, 0\,]$\,:
\be
M_g(x) &=& \frac{1}{\abs{x}} \, \int_{x}^0 dx \,
\abs{g(x)}.
\ee
The bound (\ref{ScloseNgNo}) follows from eq.~(\ref{eq:ScloseNc})
if $y_o \gg 1$ by noting that for both models
$x(y) \approx e^{-y}$, if $y\gg 1$.
\hfill $\Box$ 
\sub{Proof of theorem \ref{th:CgCurrentx}}
\label{ap:CgCurrentx}
The proofs of eqs (\ref{CgCurrentx}) and (\ref{CgCurrenty}) are
similar to that of theorem \ref{th:simpleTEI}.
From defs (\ref{CgcurrentSubx}) and (\ref{CgcurrentSuby})
the mean current transforms as a tensor:
\be
\wh{J}_g (y) &=& x'(y)\,J_g(x(y)).
\label{eq:transfocurrent}
\ee 
Using the Parseval identity and def.~(\ref{CgCharge}),
\be
Q_g^{in} &=& \intoi \sur{dk}{2k} \, \abs{\tf{g}(k)}^2,
\ee
which implies eq.~(\ref{CgChargex}).
Definitions (\ref{CgCharge}) with transformation 
(\ref{eq:transfocurrent}) imply eq.~(\ref{CgChargey}).
\hfill{$\Box$}
\sub{Proof of theorem \ref{th:ThTEI}}
\label{ap:ThTEI}
Here we only give the proof for $T_{\gb,in}^{Th}(x)$ and 
$\wh{T}_{\gb,in}^{Th}(y)$ defined by eqs
(\ref{ThdefTEIsubx}) and (\ref{ThdefTEIsuby}) respectively.
Equation (\ref{ThTEIx}) follows from  the result 
(\ref{thermalTEI}).
Differentiating eq.~(\ref{Th2PFy}) twice gives
\be
\mtin{\partial_y \ch{\phi}(y)\,
\partial_y \ch{\phi}(y+\gep)^\dagger}
&=& - \frac{\pi}{4\gb^2} \, 
\sur{x'(y+\gep)\,x'(y)}{\sinh^2\left\{\,\frac{\pi}{\gb}\,
\left[x(y+\gep)-x(y)\right]\,\right\}}\cdot
\label{ThTEIa}
\ee
Expanding this last equation at $\gep=0$,
\be
\wh{T}_{\gb,in}^{Th}(y) &=&
 \wh{T}_o(y)+ x'(y)^2 \, \frac{\pi}{12 \gb^2}\,,
\ee
which implies eq.~(\ref{ThTEIy}).
\hfill{$\Box$}
\sub{Proof of theorem \ref{th:Thbound}}
\label{ap:Thbound}
Equation (\ref{ThNa}) implies
\be
\left\vert\,\bar{N}_{\gb,in}^{Th}[f] - \bar{N}_o[f] 
\,\right\vert
&\leq& \sur{1}{2\,\gb} \,
\intii \sur{dk}{k^2} \, \abs{\tfch{f}(k)}^2.
\ee
This last term is more easily treated than the original 
expression (\ref{ThNa}) (the decreasing exponential does not 
play a crucial role here).
If $\cha{F}(x)$ and $F(y)$ are respectively defined by eqs 
(\ref{eq:defprimitivex}) and (\ref{eq:defprimitivey}), then
\beaa{rcccl}
\intii \sur{dk}{k^2} \, \abs{\tfch{f}(k)}^2
&=& \intio dx \, \abs{\cha{F}(x)}^2 
&<& \intii dy \, \abs{F(y)}^2 
\\ [3mm]
&=& \intoi \sur{dp}{p^2} \, \abs{\tf{f}(p)}^2,
\label{eq:ThboundA}
\eeaa
where it has been assumed that $x'(y) <1$, $\forall\,y\in\R$
(the proof is thus not valid for the relativistic 
black-hole model).
Equation (\ref{eq:ThboundA}) implies the bound (\ref{Thbound}).
\hfill{$\Box$}
\sub{Proof of theorem \ref{th:ThCGHSNhorizon}}
\label{ap:ThCGHSNhorizon}
Let $\xi$ be a function such that
\beaa{rcccl}
\Lim_{y_o \rightarrow +\infty} \xi(y_o) &=& 
\Lim_{y_o \rightarrow +\infty} \, [\,y_o-\xi(y_o)\,] 
&=& +\infty.
\eeaa
With hypothesis (\ref{HYPThCGHSNhorizon}) eq.~(\ref{ThNa}) 
implies that
\be
\mbox{\hspace{-10mm}}
\left\vert\,
\bar{N}_{\gb,in}^{Th}[f_{y_o}] -\bar{N}_o[f_{y_o}] 
\,\right\vert
&\leq &\sur{1}{\gb}\, \intii \sur{dk}{k^2} \, 
\abs{\fy{\tfch{f}}(k)}^2 
\nonumber \\ [3mm]
&=& \sur{1}{\gb}\,
\intii dy \ x'(y+y_0)\, \abs{F(y)}^2 
\nonumber \\ [3mm]
&\leq& \sur{1}{\gb}\,\left[\, 
\Int_{-\infty}^{-\xi(y_o)} dy \, \abs{F(y)}^2 +
\sur{1}{1+e^{y_o-\xi(y_o)}}\, \int^{\infty}_{-\xi(y_o)} dy \, 
\abs{F(y)}^2 \, \right] 
\nonumber \\ [3mm]
&\leq& \sur{1}{\gb}\, \left[\,
\sur{C^2}{\ga^2 \, (2\ga-1)\ \xi(y_o)^{2\ga-1}}
+ e^{\xi(y_o)-y_o} \, \intoi \sur{dp}{p^2} \,\abs{\tf{f}(p)}^2
\, \right],
\label{eq:ThCGHSNhorizonA}
\ee
if $\xi(y_o)\geq L$.
With $\xi(y_o)= \frac{1}{2}\,y_o +L$,
eq.~(\ref{eq:ThCGHSNhorizonA})  implies the bound 
(\ref{ThCGHSNhorizon}).
\end{document}